\documentclass[%
reprint,
superscriptaddress,
 amsmath,amssymb,
 aps,
]{revtex4-2}
\usepackage{multirow} 
\usepackage{graphicx}
\usepackage{dcolumn}
\usepackage{bm}
\usepackage{float}
\usepackage[colorlinks,
            urlcolor=blue,
            linkcolor=blue,
           anchorcolor=blue,
            citecolor=blue]{hyperref}

\begin{document}

\preprint{APS/123-QED}

\title{Acoustic circular dichroism in a three-dimensional chiral metamaterial}

\author{Qing Tong}
\affiliation{Department of Physics, City University of Hong Kong, Tat Chee Avenue, Kowloon, Hong Kong, China}%
\author{Jensen Li}
\affiliation{Department of Physics, The Hong Kong University of Science and Technology, Kowloon, Hong Kong, China}
\author{Shubo Wang}\email{shubwang@cityu.edu.hk}
\affiliation{Department of Physics, City University of Hong Kong, Tat Chee Avenue, Kowloon, Hong Kong, China}


\date{\today}
            
\begin{abstract}
Circular dichroism (CD) is an intriguing chiroptical phenomenon associated with the interaction of chiral structures with circularly polarized lights. Although the CD effect has been extensively studied in optics, it has not yet been demonstrated in acoustic systems. Here, we demonstrate the acoustic CD effect in a three-dimensional chiral metamaterial supporting circularly polarized transverse sound. We find that the effect is negligible in the lossy metamaterial possessing $C_4$ rotational symmetry but can be strongly enhanced in the $C_2$-symmetric system with inhomogeneous loss. The phenomena can be understood based on the properties of the metamaterial’s complex band structure and the quality factors of its eigenmodes. We show that the enhanced CD in the $C_2$-symmetric system is attributed to the polarization bandgaps and the non-Hermitian exceptional points appearing near the Brillouin-zone center and boundaries. The results contribute to the understanding of chiral sound-matter interactions and can find applications in acoustic sensing of chiral structures and sound manipulations based on its vector properties. 
\end{abstract}

\maketitle


\section{\label{sec:level1}INTRODUCTION}

Chiral structures have novel properties deriving from the mirror-symmetry breaking \cite{2017chiral, ref3} and are extensively employed to realize polarization conversion \cite{2009gold, wu2011}, unusual optical forces  \cite{ref12, ref13, ref14}, and synthetic gauge fields\cite{2013photonic}. The interaction between chiral structures and chiral light, i.e., light carrying spin and/or orbital angular momentum (OAM), can give rise to circular dichroism (CD) \cite{ref1, ref3, ref19} and helical (or vortical) dichroism \cite{ref21, ref22}, corresponding to the differential absorption of lights with opposite chirality.  The CD effect has been investigated in various optical structures, ranging from bilayer chiral structures \cite{ref26, ref28}, nonplanar three-dimensional (3D) chiral structures \cite{ref7}, to gyroid structures \cite{ref29, ref31}. Recent research has uncovered the subtle relations between CD and the Ohmic dissipation of meta-atoms \cite{ref32} as well as the bound states in the continuum \cite{shi2022}, enabling a profound understanding of chiral light-matter interactions. The CD effect has been widely applied to analyze molecular structures \cite{ref34,ref37} and to achieve chiral discrimination \cite{2018circular,2022chiral}. 

It is well known that sound can carry OAM in the form of vortices \cite{ref40, ref41, ref42}. The acoustic OAM can induce chiral sound-matter interactions and give rise to intriguing phenomena such as acoustic geometric phases \cite{ref46} and the acoustic orbital Hall effect \cite{ref54, ref55}. The chiral sound-matter interactions can enable rich manipulations of sound vortices, including asymmetric transmission/reflection \cite{ref53}, reversal of orbital angular momentum \cite{ref56}, and acoustic helical dichroism \cite{ref45}.  In addition, the chiral sound-matter interactions can be applied to manipulate matter, leading to acoustic levitation \cite{ref47, ref48}, acoustic tweezers \cite{ref44, ref49, ref50}, and acoustic torque \cite{ref51, ref52}. While airborne sound is a longitudinal wave, it was shown that inhomogeneous sound fields can carry nonzero acoustic spin density characterized by rotating velocity vector fields \cite{ ref58, ref60}. Remarkably, spin-1 transverse sound can emerge in a micropolar metamaterial supporting synthetic shear forces in air \cite{ref61}. In contrast to the conventional longitudinal sound, the transverse sound carries full vector properties similar to electromagnetic waves. In particular, it can carry both spin and OAM with intriguing spin-orbit interactions. Exploration of this new type of sound and its counter-intuitive properties can generate new functionalities for acoustic applications. 

Here, for the first time, we demonstrate the acoustic CD effect in a 3D chiral metamaterial that supports circularly polarized transverse sound.  Using full-wave numerical simulations, we calculate the absorption of left-handed circularly polarized (LCP) and right-handed circularly polarized (RCP) sound in the lossy metamaterial. We find that the CD effect strongly depends on the rotational symmetry of the metamaterial. The CD effect is negligible  in the metamaterial with homogeneous loss satisfying the $C_4$ rotational symmetry. In contrast, it is strongly enhanced in the metamaterial with inhomogeneous loss satisfying the $C_2$ rotational symmetry. By studying the complex band structure of the metamaterial and the quality factors of its eigenmodes, we find that these properties originate from the polarization bandgaps and the non-Hermitian exceptional points (EPs) of the metamaterial.

We organize the article as follows. In Section \ref{sec:level2}, we introduce the acoustic chiral metamaterial and discuss its eigenmode properties. In Section \ref{sec:level3}, we show the numerical results for the absorption of the LCP and RCP sound in two types of lossy metamaterial obeying the $C_4$ and $C_2$ rotational symmetry, respectively. Section \ref{sec:level4} presents the complex band structures of the lossy metamaterials, where we discuss the polarization bandgaps and the EPs to understand the CD effect. We draw the conclusion in Section \ref{sec:level5}. 

\section{\label{sec:level2}THE CHIRAL METAMATERIAL}

\begin{figure}[h!]
\centering
\includegraphics[scale=1.2]{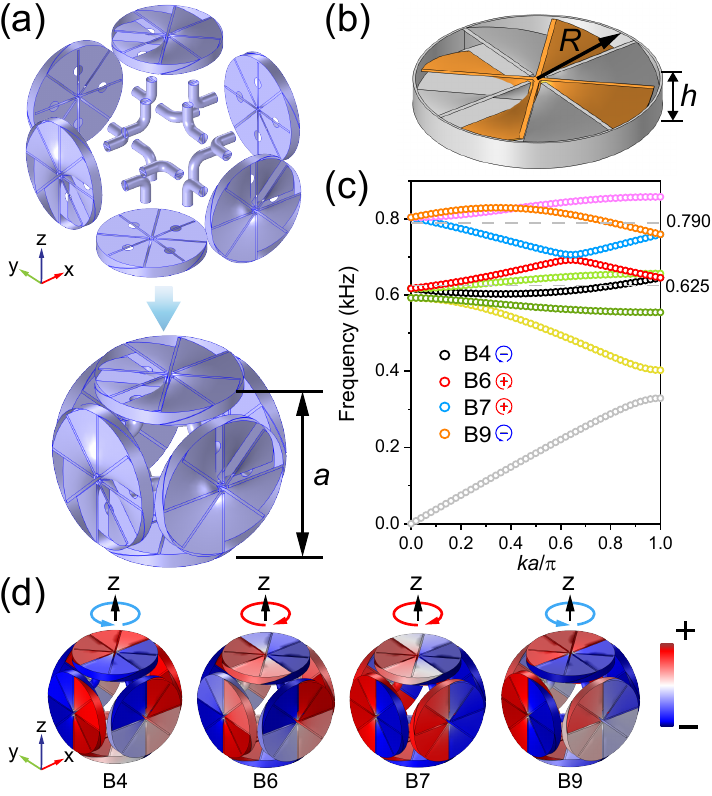}
\caption{ (a) Unit cell of the 3D acoustic metamaterial. (b) Internal structure of the resonators. The geometry parameters are $R=5$ cm, $h=1$cm, $a=12.1$ cm. (c) Band structure of the metamaterial. B4, B6, B7, and B9 denote the lowest four bands with circularly polarized eigenstates. (d) Pressure fields at $ka/\pi=0.2$ for B4, B6, B7, and B9. The blue and red arrows denote the circulating direction of the eigen pressure fields for the LCP and RCP states, respectively. } \label{fig1}
\end{figure}

We consider a 3D metamaterial with the cubical unit cell shown in Figure \ref{fig1}(a). The unit cell comprises three chiral resonators mutually connected by tubes, and it obeys the $C_4$ rotational symmetry with respect to the $x, y,$ and $z$ axes. Figure \ref{fig1}(b) shows a half of the chiral resonator, where the radius is $R=5$ cm, and the height is $h=1$ cm. The orange blades are connected to the center post, and the gray blades are connected to the outer shell of the resonator. We assume that air is filled inside the resonator and all air-material interfaces are hard boundaries. In the considered range of frequencies, each resonator support subwavelength resonances. These resonances endow the metamaterial with intriguing macroscopic acoustic properties. 

We first calculate the band structure of the metamaterial by using a finite-element package COMSOL Multiphysics. The result is shown in Figure \ref{fig1}(c) for the wavevector in $z$ direction. The 4th to 9th bands (counting from bottom) correspond to the eigenmodes dominated by the dipole resonances of the chiral resonators. Here, we focus on the bands labeled as B4, B6, B7, and B9. The eigenmodes of these bands are circularly polarized transverse sound \cite{ref61}, and their pressure fields are shown in Fig. \ref{fig1}(d) for $ka/\pi=0.2$. As seen, the fields represent acoustic dipoles oscillating in a direction perpendicular to the axis of the resonators.  The arrowed circles denote the rotation direction of the eigen fields. The red and blue arrowed circles correspond to the right circularly polarized (RCP) states and left circularly polarized (LCP) states, respectively. The velocity field averaged over the unit cell also circulates in the same direction. The collective motion of the acoustic dipoles gives rise to circularly-polarized transverse sound macroscopically. The transverse sound of the bands B4 and B6 (B7 and B9) have opposite handedness and carry opposite spin angular momentum.  

\begin{figure}[h!]
\centering
\includegraphics[scale=1.2]{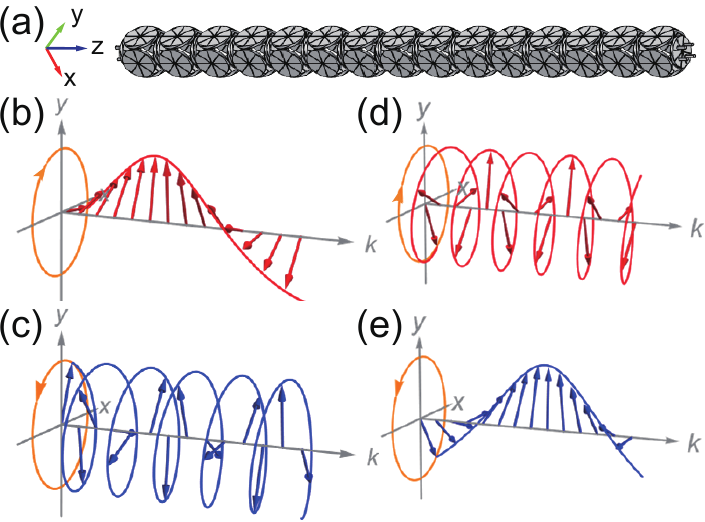}
\caption{(a) The metamaterial with 15 units in $z$ direction and periodic in $x$ and $y$ directions. The transverse sound is excited by ports on the left end of the metamaterial. (b-e) Averaged velocity vectors in the metamaterial at the frequency $f=0.625$ kHz (b,c) and $0.79$ kHz (d,e). The yellow circles with the arrow show the circulating direction of the velocity field on the $xy-$plane.} \label{fig2}
\end{figure}

To demonstrate the emergence of circularly polarized transverse sound, we consider the metamaterial with 15 unit cells in $z$ direction, as shown in Fig. \ref{fig2}(a). Periodic boundary conditions are applied in $x$ and $y$ directions. To excite the system, we set four input ports at the four tubes on the left side of the unit in Fig. \ref{fig2}(a) with the phases 0, $0.5\pi$, $\pi$, and $1.5\pi$, respectively. The azimuthal gradient of the phase decides the handedness (LCP or RCP) of the excited circularly-polarized transverse sound. In addition, we set another four output ports on the right side of the metamaterial to determine the transmission. We consider two frequencies, 0.625 kHz and 0.79 kHz corresponding to the frequencies marked by the dashed lines in Fig. \ref{fig1}(c), where four eigenstates (two are of LCP and two are of RCP) can be excited. To visualize the transverse sounds, we average the velocity field over each unit cell and plot the averaged velocity vectors (denoted by the \textcolor{black}{red and blue} arrows) for the 15 unit cells in Fig. \ref{fig2}(b)-(e). We observe that the velocity vectors indeed rotate in the $xy$ plane, corresponding to spin angular momentum in the longitudinal direction (i.e., z direction). The arrowed yellow circles denote the temporal evolution of the velocity vectors, with the arrows indicating the circulation direction. At either frequency, the two transverse sounds carry opposite spins, allowing the exploration of their different absorption inside the metamaterial. 

\section{\label{sec:level3}ACOUSTIC CIRCULAR DICHROISM}

\begin{figure}[tb]
\centering
\includegraphics[scale=1.2]{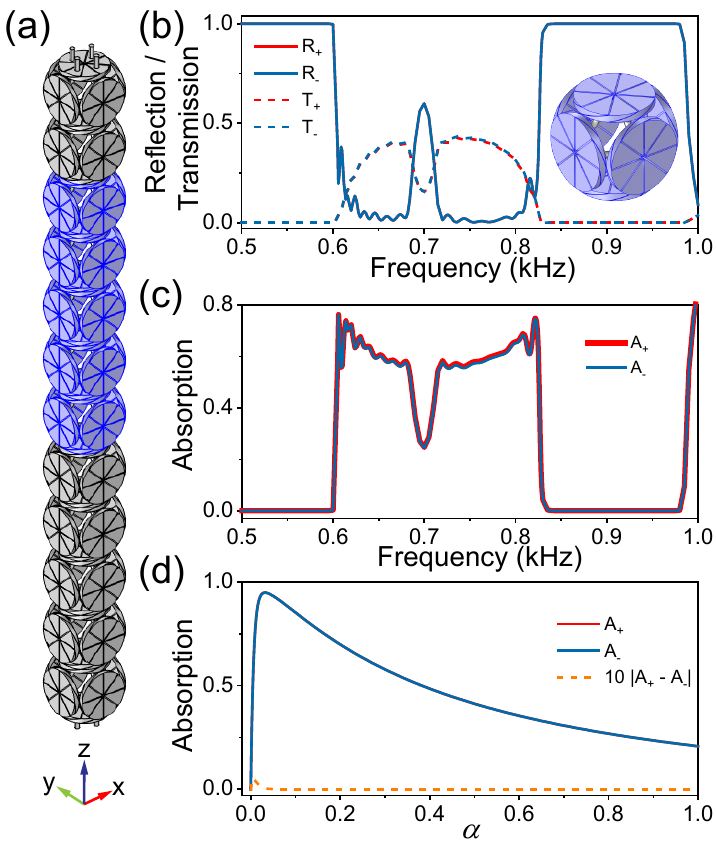}
\caption{(a) The 3-layer metamaterial with loss uniformly added to the middle layer (colored in yellow). (b) Reflection, transmission, and (c) absorption as a function of frequency for the RCP (+) and LCP (-) sounds with $\alpha=0.006$. The inset in (b) shows the regions containing loss (blue colored), which satisfies the $C_4$ symmetry. (d) Absorption of the LCP and RCP sounds as a function of the loss $\alpha$ at {\color{black} 0.65 kHz}. In (c) and (d), the absorption difference ($\Delta A$) is multiplied by ten for easy visualization.}  \label{fig3}
\end{figure}

We now consider the metamaterial with loss to investigate the CD effect of the transverse sound. As shown in Fig. \ref{fig3}(a), we employ a three-layer sandwich structure: the bottom and top layers are lossless, while the middle layer (highlighted in  blue) contains loss. The loss is introduced into the unit cells by adding an imaginary part to the sound speed $v(1+i\alpha)$ with $\alpha$ characterizing the loss strength. We apply ports to excite the system (same as in Fig. \ref{fig2}) from the bottom of the lattice. The excited transverse sound propagates through the middle lossy layer and is measured in the top layer to determine its transmission. This directly maps to the usual configuration of optical CD, where an optical structure is sandwiched by air and reflection/transmission is measured in air. We apply COMSOL to simulate the system and calculate the reflection ($R_{\pm}$) and transmission ($T_{\pm}$) of the incident sound. The absorptions can then be determined as:

\begin{align}
A_{\pm}=1-R_{\pm}-T_{\pm}, \label{eq:1}\\
\Delta A=\left|A_{+}-A_{-}\right|, \label{eq:2}
\end{align}

\noindent where ``$+$'' (``$-$'') denotes the RCP (LCP) state and $\Delta A$ is the differential absorption of the RCP and LCP sounds.

We first consider the case with loss uniformly added to all the chiral resonators in the middle blue-colored layer. Figure \ref{fig3}(b) shows the numerical results of the reflections (solid lines) and transmissions (dashed lines). The inset shows the unit cell with the blue-colored regions containing loss $\alpha=0.006$ \textcolor{black}{, which has $C_4$ rotational symmetry}. As seen, the reflection $R_+$ is nearly identical to $R_-$, and there is a tiny difference between the transmissions $T_+$ and $T_-$. The absorptions calculated using Eq.(\ref{eq:1}) are shown in Fig. \ref{fig3}(c) as the solid blue and red lines. We notice that the absorptions of LCP and RCP sounds are \textcolor{black}{almost equal}.  \textcolor{black}{As a result}, the CD effect is {\color{black} negligible} in this case with homogeneous loss. Figure \ref{fig3}(d) shows the absorption of the LCP and RCP sound as a function of the loss strength $\alpha$ at $f=0.65$ kHz. We notice that the absorption $A_+$ and $A_-$ are almost identical. The differential absorption $\Delta A$ is less than 0.01 with the maximum appears at $\alpha=0.006$, as shown by the orange dashed line which has been multiplied by a factor of 10.  

\begin{figure}[h!tpb]
\centering
\includegraphics[scale=1.2]{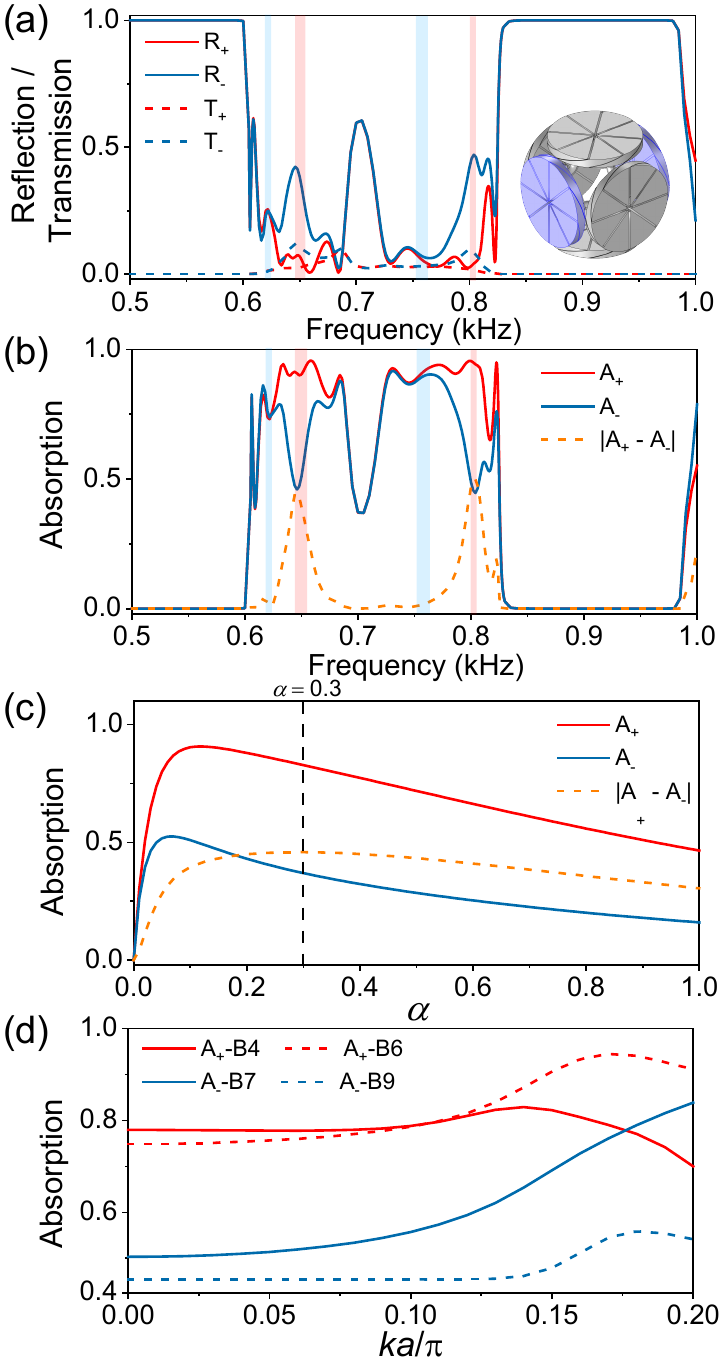}
\caption{(a) Reflection, transmission, and (b) absorption of the RCP (+) and LCP (-) sounds for $\alpha=0.1$. The inset in (a) shows the regions with loss (blue colored), which satisfies the $C_2$ symmetry. (c) Absorption of the LCP and RCP sounds as a function of the loss parameter $\alpha$ at 0.65 kHz. (d) Absorption of the LCP and RCP sounds as a function of the normalized wavenumber $ka/\pi$ for the bands B4, B6, B7, and B9. } \label{fig4}
\end{figure}

The optical CD effect strongly depends on the rotational symmetry of the structures \cite{2010twisted, 2015circular}. To explore this symmetry dependence for acoustic CD, we break the $C_4$ symmetry of the metamaterial by selectively adding loss to the unit cells. As shown in the inset of Fig. \ref{fig4}(a), we only add loss to the side resonator (i.e., two opposing half resonators highlighted in blue) with the center axis in $x$-direction, reducing the symmetry of the metamaterial from $C_4$ to $C_2$. We numerically calculated the transmission and reflection of this $C_2$ system, and the results are shown in Fig. \ref{fig4}(a) for loss $\alpha=0.1$. We observe large differences in the transmissions and reflections of the LCP and RCP sound in the frequency range {\color{black}[0.600 kHz, 0.825 kHz]}, corresponding to the considered bands in Fig. \ref{fig1}(c). The absorptions calculated with Eq. (\ref{eq:1}) are shown in Fig. \ref{fig4}(b). As noticed, there is a significant difference between the absorption of RCP sound ($A_+$) and the absorption of RCP sound ($A_-$). The differential absorption $\Delta A$ (denoted by the orange dashed line) is much larger than the case of Fig. \ref{fig3} and has two local maxima of about 0.5 appearing at 0.65 kHz and 0.80 kHz. This demonstrates the strong CD phenomena in the $C_2$ metamaterial. We also investigate the dependence of the CD on the loss magnitude $\alpha$, and the results are shown in Fig. \ref{fig4}(c) for $f = 0.65$ kHz. The trends of $A_+$ and $A_-$ are similar to those of the $C_4$ system, but the absorption difference $\Delta A$ (denoted by the dashed orange line) is much larger with a maximum value of 0.46 at $\alpha=0.3$ (marked by the dashed line).  

The CD characterizes the different absorption of LCP and RCP sounds at the same frequency. We note that the LCP and RCP sounds have different wavelengths inside the chiral metamaterial at the same frequency due to their different dispersions. It is thus interesting to compare their absorption for the same wavelength (or wavenumber) inside the metamaterial. Figure \ref{fig4}(d) shows the absorption of the transverse sound corresponding to the four bands B4, B6, B7, and B9 in Fig. \ref{fig1}(c). We only consider the range of $0 \leq ka/\pi\leq 0.2$ where the effective wavelength is well defined, \textcolor{black}{and the excited state is RCP for band B4 and LCP for band B7 due to their negative group velocity in this range}. As seen, the RCP sounds (corresponding to the solid and dashed red lines) generally have larger absorption compared with the LCP sounds (corresponding to the solid and dashed blue lines). This indicates that a strong CD effect also happens to the circularly polarized transverse sounds with the same wavelength (but not necessarily the same frequency).    

\begin{figure}[h!tpb]
\centering
\includegraphics[scale=1.2]{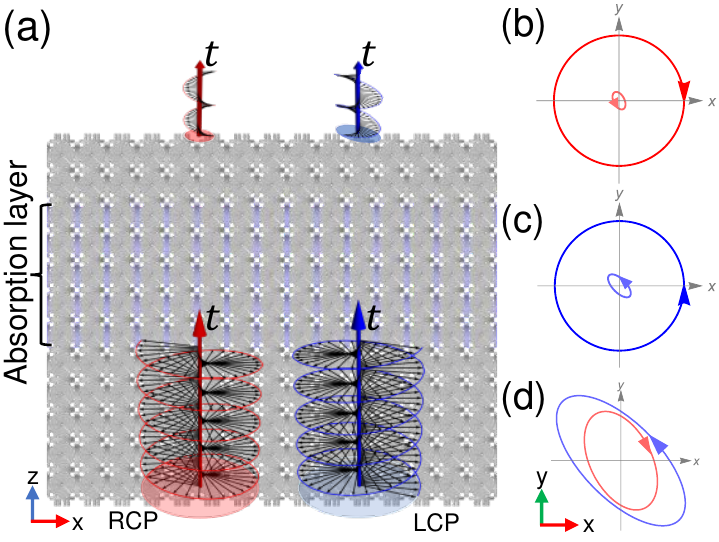}
\caption{(a) Schematics of the CD effect in the acoustic metamaterial. The red and blue helical curves denote the temporal trajectories of the velocity vectors for the RCP and LCP sounds on the incident side (bottom) and the transmission side (top). (b, c) Larger (smaller) circles denote the evolution trajectories of velocity field for the incident (transmitted) RCP and LCP sounds. The transmitted sounds are elliptically polarized. (d) A zoom-in comparison of the transmitted velocity field's trajectories under the incidence of RCP (red) and LCP (blue) sound. We set the frequency $f= 0.635$ kHz and loss $\alpha=0.1$. } \label{fig5}
\end{figure}

To intuitively understand the different absorption of LCP and RCP sounds, we then study the averaged velocity fields (averaged over one unit cell) in the 3-layer metamaterial with $C_2$ symmetry. Figure \ref{fig5}(a) shows a side view of the metamaterial, where loss is added to the middle layers consisting of 5 unit cells (the blue color marks the resonators containing loss). The red and blue helical curves in the bottom layer denote the temporal trajectories of the velocity vectors of the incident RCP and LCP sounds, respectively. The helical curves in the upper layer denote the temporal trajectories of the velocity vectors of the transmitted sounds, which are in general elliptically polarized due to the coupling between the LCP and RCP sounds in the $C_2$ absorptive layer. Figure \ref{fig5}(b) and (c) show the numerical results of the transmitted velocity fields under the incidence of the RCP and LCP sounds, respectively, for $f = 0.635$ kHz and $\alpha=0.1$. The larger arrowed circles denote the time-evolution trajectories of the incident velocity fields, while the smaller ellipses denote the time-evolution trajectories of the transmitted velocity fields. Figure \ref{fig5}(d) shows a comparison between the transmitted velocity fields under the incidence of LCP and RCP sounds (corresponding to a zoom-in of the results in Fig. \ref{fig5}(b) and (c)), which are different in both amplitude and ellipticity. Similar property also exists in the reflected fields.

\section{\label{sec:level4}COMPLEX BAND STRUCTURE AND EXCEPTIONAL POINTS}    

\begin{figure}[h!tpb]
\centering
\includegraphics[scale=1.15]{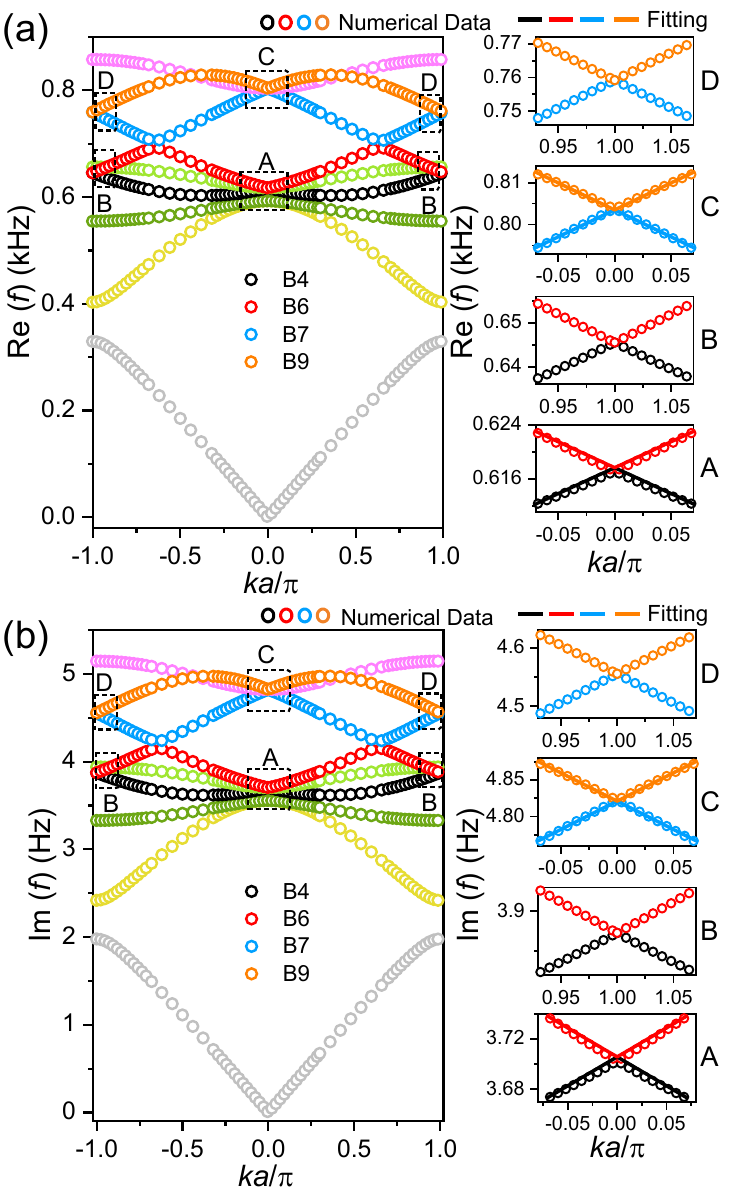}
\caption{The real (a) and imaginary (b) parts of the complex band structure for the $C_4$ system at $\alpha=0.006$. Insets on the right side show the zoom-ins of the bands near the zone center and boundaries, corresponding to the dashed rectangles in (a) and (b). The solid lines in the insets denote the analytical fitting results. } \label{fig6}
\end{figure}

We investigate the complex band structures of the systems to uncover the origins of the acoustic CD and the different properties of the $C_2$ and $C_4$ systems. Figure \ref{fig6}(a) and (b) show the real and imaginary parts of the complex band structure for the $C_4$ system with loss $\alpha=0.006$ (corresponding to the case of Fig. \ref{fig3}). The imaginary parts take positive values due to the time convention $e^{\text{i}\omega t}$ adopted in COMSOL. The insets (labeled as A, B, C, and D) on the right side show the zoom-ins of the bands enclosed by the dashed rectangles. The insets A and B depict the bands near the Brillouin zone centers and boundaries, respectively, for B4 and B6. Likewise, insets C and D show the bands closed to the Brillouin zone centers and boundaries, respectively, for B7 and B9. We notice that the band degeneracies are not affected by the loss due to the protection of the $C_4$ symmetry.

\begin{figure}[h!tpb]
\centering
\includegraphics[scale=1.15]{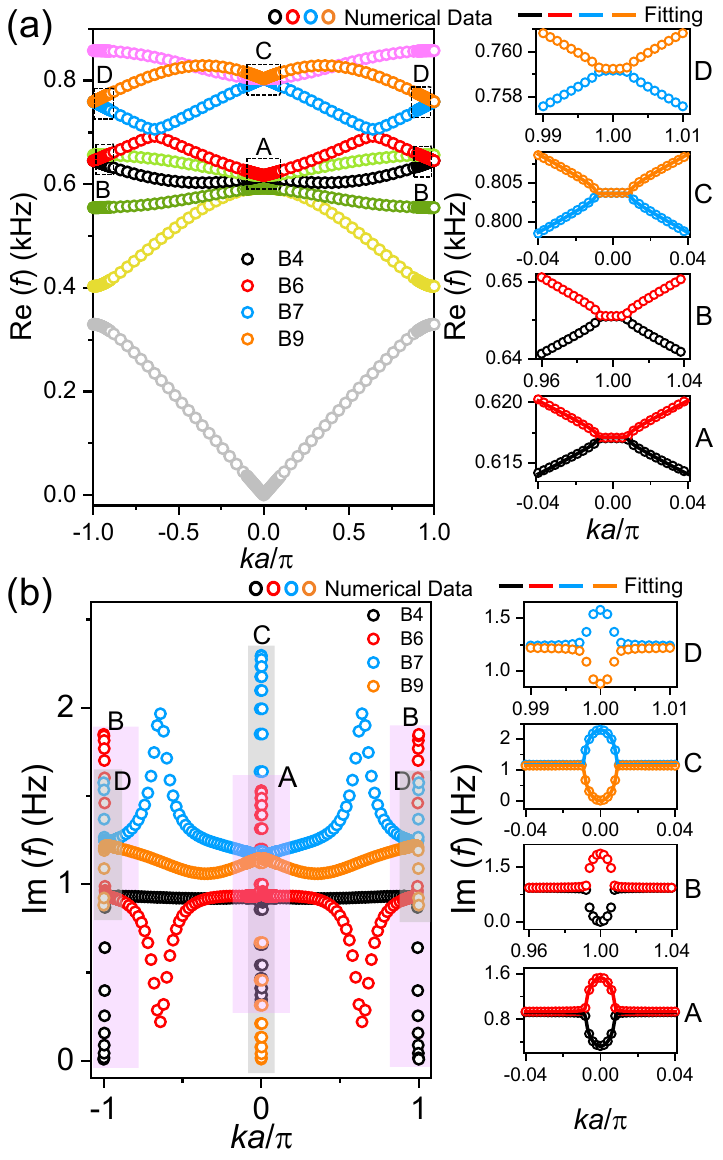}
\caption{The real (a) and imaginary (b) parts of complex band structure for the $C_2$ system at $\alpha=0.006$. The insets on the right side show the zoom-ins of the bands near the zone center and boundaries, corresponding to the dashed rectangles in (a) and shaded rectangles in (b). The solid lines in the insets denote the analytical fitting results.} \label{fig7}
\end{figure}

Figure \ref{fig7} shows the complex band structure for the $C_2$ system with the same loss $\alpha=0.006$ for comparison with the $C_4$ system. Interestingly, at the Brillouin zone center and boundaries, the real parts of the bands remain degenerate in a finite range of $k$ values while the imaginary parts bifurcate in the same range, as shown in the insets on the right side. This indicates the emergence of non-Hermitian exceptional points \cite{ref63, ref64, ref65}. Obviously, these EPs derive from the diabolic points of the original lossless system in Fig. \ref{fig1} (c). While the phenomena here is similar to the EPs spawn from Dirac points in two-dimensional photonic crystals \cite{ref63}, the underlying physical mechanism is different. The emergence of these EPs is attributed to the coupling and loss difference of the LCP and RCP transverse dipole modes induced by the breaking of $C_4$ symmetry. We will elaborate on this point with an analytical model later.  

\begin{figure}[tb]
\centering
\includegraphics[scale=1.15]{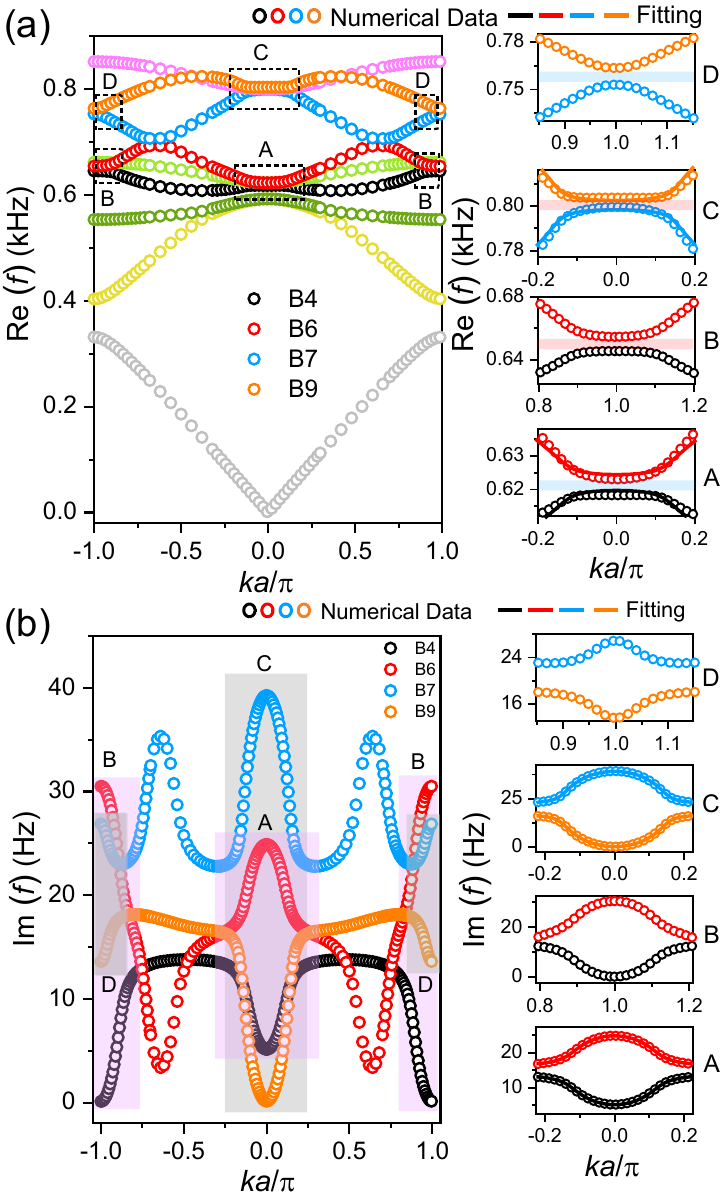}
\caption{The real (a) and imaginary (b) parts of complex band structures for the $C_2$ system at $\alpha=0.1$. The right insets show the zoom-ins of the bands near the Brillouin zone centers and boundaries, corresponding to the dashed rectangles in (a) and shaded rectangles in (b). The ribbons in the insets denote the partial bandgaps. The solid lines in the insets denote the analytical fitting results.} \label{fig8}
\end{figure}

Figure \ref{fig8}(a) and (b) show the complex band structure for the $C_2$ system with a larger loss $\alpha=0.1$, corresponding to the case of Fig. \ref{fig4}(a) and (b) with a much stronger CD effect. We notice that the EP features remain at the center and boundaries of the Brillouin zone. At the same time, small partial gaps appear at the frequencies of the EPs, as marked by the blue and red ribbons in the insets of Fig. \ref{fig8}(a). At the frequencies of the blue-ribbon (red-ribbon) region, only LCP (RCP) sound can propagate through the metamaterial \cite{ref29, ref65}. Thus, at the frequencies $f = 0.62$ kHz and $f = 0.76$ kHz, the RCP sound cannot propagate through the metamaterial. Similarly, at the frequencies of $f = 0.65$ kHz and $f = 0.80$ kHz, the LCP sound cannot propagate through the metamaterial. However, this does not necessarily indicate a large difference in the reflection of LCP and RCP sounds at these frequencies due to the non-Hermitian nature of the metamaterial. Whether strong reflection will appear at the partial polarization gaps depends on the damping of the corresponding eigenmodes. In the following, we will show that the eigenmodes’ damping property strongly affects the reflection and the CD effect. 

\begin{figure}[h!tpb]
\centering
\includegraphics[scale=1.2]{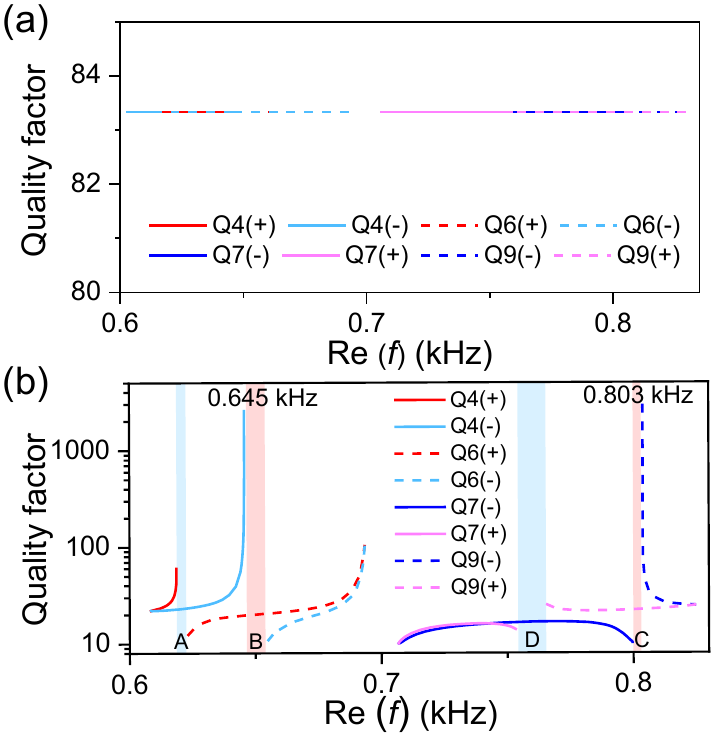}
\caption{The quality factor $Q$ of the eigenmodes in the (a) $C_4$ and (b) $C_2$ systems, corresponding the cases of Fig. 6 and Fig. 8, respectively. The blue and red ribbons denote polarization bandgaps.} \label{fig9}
\end{figure}

To understand the damping of the eigenmodes, we investigate the modes’ quality factor $Q$ corresponding to the bands B4, B6, B7, and B9, for both the $C_4$ and $C_2$ systems. The quality factor is calculated as $ Q=\frac{\rm{Re}(\textit{f})}{2 \rm{Im}(\textit{f})}$ \cite{ref66}. The results are shown in Fig. \ref{fig9}(a) and (b) as a function of the real part of the eigenfrequency. We note that the eigenmode of each band can be either LCP or RCP, depending on the sign of its group velocity with respect to the phase velocity. Consequently, the quality factor $Q$ of each band can be divided into two parts for the LCP (“$-$”) and RCP (“$+$”) states, respectively. As shown in Fig. \ref{fig9}(a), all eigenmodes of the $C_4$ system have approximately the same quality factor. This explains the {\color{black} negligible} CD effect in the $C_4$ system with homogenous loss. In contrast, the quality factors of the LCP and RCP modes in the $C_2$ system have a large difference, particularly in the vicinity of the polarization bandgaps marked by the red ribbons. The large difference in quality factor indicates a large difference in the damping of LCP and RCP modes and thus explains the strong CD effect near the polarization bandgaps, in agreement with the numerical results in Fig. \ref{fig4}(b). For the LCP and RCP modes near the blue-ribbon band gaps, their quality factors are much smaller than the modes near the red-ribbon bands, and the difference of their quality factors are also much smaller. Therefore, both LCP and RCP sounds at the frequencies of the blue-ribbon region are strongly absorbed, and their reflections are  small, leading to a weak CD effect, as confirmed by the numerical results in Fig. \ref{fig4}(b). 

To understand the emergence of the EPs in the $C_2$ system, we exploit an effective Hamiltonian to describe the coupling of the LCP and RCP modes near the Brillouin zone center \cite{ref63, ref64, fanPRL, JL2015PRA}. As for the $C_4$ system with homogenous loss, the effective Hamiltonian can be expressed as:
  
\begin{equation}       
H_{C_4}=\left(                 
  \begin{array}{cc}   
    \omega_0-\textit{i} \gamma & (v_R+\textit{i} v_I)k\\ 
    (v_R+\textit{i} v_I)k & \omega_0-\textit{i} \gamma \\ 
  \end{array}
\right) \label{eq:3}
\end{equation}

\noindent which has the complex eigenvalues:

\begin{equation}       
\omega=\omega_0 -\textit{i} \gamma \pm k (v_R+\textit{i} v_I).
\label{eq:4}
\end{equation}

\noindent Here, $\omega_0$ is the eigenfrequency at $k=0$, where the LCP and RCP modes are degenerate; $v_R$ and $v_I$ are the real and the imaginary parts of the complex group velocities, respectively; $\gamma$ denotes the loss. 

In the $C_2$ system, loss is selectively added to only one resonator in each unit cell. The breaking of $C_4$ symmetry opens a gap at $k = 0$, which can be characterized by a perturbation term $\delta/2$ in the Hamiltonian. The LCP and RCP modes at $k = 0$ now have different loss $\gamma_1$ and $\gamma_2$ ($\gamma_1\ne\gamma_2$):

\begin{equation}       
H_{C_2}=\left(                 
  \begin{array}{cc}   
    \omega_0-\textit{i} \gamma_1 + \frac{\delta}{2} & (v_R+\textit{i} v_I)k\\ 
    (v_R+\textit{i} v_I)k & \omega_0-\textit{i}\gamma_2-\frac{\delta}{2} \\ 
  \end{array}
\right) , \label{eq:5}
\end{equation}

\noindent which has the complex eigenvalues:

\begin{equation}      
\begin{split}
\omega=&\omega_0 - \textit{i} \frac{(\gamma_1 +\gamma_2)}{2} \\
&\pm \frac{1}{2} \sqrt{[\delta- \textit{i}(\gamma_1 - \gamma_2)]^2 - 4k^2 (v_I - \textit{i} v_R)^2}. 
\label{eq:6}
\end{split}
\end{equation}

These analytical expressions of the complex eigenvalues in Eq.(\ref{eq:4}) and (\ref{eq:6}) are employed to fit the numerical results for both the real and the imaginary parts. The fitting results are shown as the solid lines in the insets of Figs. \ref{fig6}-\ref{fig8}, accordingly. We notice good quantitative agreements between the analytical and numerical results. In particular, the effective Hamiltonian correctly captures the EP features in the $C_2$ systems. {\color{black} The fitting parameters for both $C_4$ and $C_2$ systems with different losses are summarized in Table \ref{tab1}.} We note that the mode damping parameters $\gamma_{1,2}$ take negative values due to the time convention $e^{\text{i}\omega t}$ adopted in COMSOL. 

\begin{table}[hbtp]

\caption{\centering Fitting parameters for $C_4$ and $C_2$ systems}

\renewcommand\arraystretch{1.3}
\small
\centering
\scalebox{0.9}
{
\begin{tabular}{ccccccccc}
\toprule 
System  & $\alpha$    & $\omega_0$   & \multicolumn{2}{c}{$\gamma$}     & $v_R$  & $v_I$ &  & Inset\\ \hline
\multirow{2}{3mm}{$C_4$} & \multirow{2}{9mm}{0.006} & 617.61 & \multicolumn{2}{c}{-3.71}  & 2.96 & 0.018 & & A (Fig. \ref{fig6}) \\ 
     &           & 803.5  & \multicolumn{2}{c}{-4.82}     & 4.97 & 0.03   &       & C (Fig. \ref{fig6}) \\ \hline
\multirow{5}{3mm}{$C_2$} & $\alpha$ & $\omega_0$  & $\gamma_1$ & $\gamma_2$ & $v_R$ & $v_I$ &$\delta$ & Inset\\ \cline{2-9}
            & \multirow{2}{9mm}{0.006} & 617.12 & -1.53  & -0.33  & 2.96 & 0.0011 & 0.036  & A (Fig. \ref{fig7}) \\ 
            &                        & 803.66 & -0.01  & -2.30  & 4.99 & -0.013 & 0.0075 & C (Fig. \ref{fig7}) \\ \cline{2-9} 
            & \multirow{2}{4mm}{0.1}   & 622.12 & -24.87 & -5.13  & 2.98 & 0.03   & 4.77   & A (Fig. \ref{fig8}) \\ 
            &                        & 800.40 & -0.22  & -39.12 & 4.98 & -0.21  & 4.22   & C (Fig. \ref{fig8}) \\ \toprule

\end{tabular}}
\label{tab1}
\end{table}





The above effective Hamiltonians well explain the emergence of the EPs and the enhancement of CD by the EPs. In the $C_4$ system, the LCP mode of the B4 band and the RCP mode of the B6 band are orthogonal at $k = 0$ with vanished coupling. The damping of the LCP and RCP modes at the same excitation frequency are approximately equal due to homogeneous loss added to all resonators of the unit cell. Thus, their quality factors are almost equal (corresponding to the results in Fig. \ref{fig9}(a)). In the $C_2$ system, the inhomogeneous loss breaks the $C_4$ rotational symmetry and induces coupling between the original LCP and RCP modes at $k = 0$, which gives rise to the polarization bandgaps. In addition, the two modes have different dampings due to the inhomogeneous material loss. These together give rise to the EPs and the bifurcation of the imaginary parts of the eigenfrequencies, leading to enlarged damping contrast of the LCP and RCP modes at the same excitation frequency and thus larger difference in their quality factors (corresponding to the results in Fig. \ref{fig9}(b)). Therefore, the strong CD effect in the $C_2$ system is attributed to both the polarization bandgaps and the EPs.

\section{\label{sec:level5} CONCLUSION }

In conclusion, we demonstrate the acoustic CD effect in a 3D chiral metamaterial supporting circularly polarized transverse sound. We have investigated the effect in two types of systems with $C_4$ and $C_2$ rotational symmetry, respectively. In the $C_4$ system with loss homogeneously added to all resonators of the  unit cell, we observe a  negligible acoustic CD effect. On the other hand, by selectively adding loss to part of the unit cell, reducing the system’s rotational symmetry from $C_4$ to $C_2$, the CD effect is enhanced strongly. With analysis of their complex band structures and quality factors of the eigenmodes, we uncover that the strong acoustic CD in the $C_2$ system is attributed to polarization bandgaps and the emergence of non-Hermitian EPs. The polarization bandgaps induce selective transmission and absorption of the circularly polarized transverse sound with a particular handedness. The EPs give rise to bifurcations of the imaginary parts of the eigen frequencies. These together enhance the CD effect in the $C_2$ system. It will be interesting to experimentally demonstrate the discussed phenomena. The metamaterial structures can be fabricated using 3D printing. Loss can be introduced into the structures by adding sponges. The transverse sound can be excited by using an array of speakers, and the reflection/refraction can be measured with a microphone. The acoustic CD effect can find applications in sound manipulations based on its vector degrees of freedom and in acoustic sensing of chiral structures. The results contribute to the understanding of chiral sound-matter interactions in metamaterials and phononic crystals.

\begin{acknowledgments}
The work described in this paper was supported by grants from the Research Grants Council of the Hong Kong Special Administrative Region, China (No. CityU 21302018 and No. C6013-18G).   
\end{acknowledgments}


\bibliography{MyCollection}

\begin{thebibliography}{50}%
\makeatletter
\providecommand \@ifxundefined [1]{%
 \@ifx{#1\undefined}
}%
\providecommand \@ifnum [1]{%
 \ifnum #1\expandafter \@firstoftwo
 \else \expandafter \@secondoftwo
 \fi
}%
\providecommand \@ifx [1]{%
 \ifx #1\expandafter \@firstoftwo
 \else \expandafter \@secondoftwo
 \fi
}%
\providecommand \natexlab [1]{#1}%
\providecommand \enquote  [1]{``#1''}%
\providecommand \bibnamefont  [1]{#1}%
\providecommand \bibfnamefont [1]{#1}%
\providecommand \citenamefont [1]{#1}%
\providecommand \href@noop [0]{\@secondoftwo}%
\providecommand \href [0]{\begingroup \@sanitize@url \@href}%
\providecommand \@href[1]{\@@startlink{#1}\@@href}%
\providecommand \@@href[1]{\endgroup#1\@@endlink}%
\providecommand \@sanitize@url [0]{\catcode `\\12\catcode `\$12\catcode
  `\&12\catcode `\#12\catcode `\^12\catcode `\_12\catcode `\%12\relax}%
\providecommand \@@startlink[1]{}%
\providecommand \@@endlink[0]{}%
\providecommand \url  [0]{\begingroup\@sanitize@url \@url }%
\providecommand \@url [1]{\endgroup\@href {#1}{\urlprefix }}%
\providecommand \urlprefix  [0]{URL }%
\providecommand \Eprint [0]{\href }%
\providecommand \doibase [0]{https://doi.org/}%
\providecommand \selectlanguage [0]{\@gobble}%
\providecommand \bibinfo  [0]{\@secondoftwo}%
\providecommand \bibfield  [0]{\@secondoftwo}%
\providecommand \translation [1]{[#1]}%
\providecommand \BibitemOpen [0]{}%
\providecommand \bibitemStop [0]{}%
\providecommand \bibitemNoStop [0]{.\EOS\space}%
\providecommand \EOS [0]{\spacefactor3000\relax}%
\providecommand \BibitemShut  [1]{\csname bibitem#1\endcsname}%
\let\auto@bib@innerbib\@empty
\bibitem [{\citenamefont {Hentschel}\ \emph {et~al.}(2017)\citenamefont
  {Hentschel}, \citenamefont {Sch{\"a}ferling}, \citenamefont {Duan},
  \citenamefont {Giessen},\ and\ \citenamefont {Liu}}]{2017chiral}%
  \BibitemOpen
  \bibfield  {author} {\bibinfo {author} {\bibfnamefont {M.}~\bibnamefont
  {Hentschel}}, \bibinfo {author} {\bibfnamefont {M.}~\bibnamefont
  {Sch{\"a}ferling}}, \bibinfo {author} {\bibfnamefont {X.}~\bibnamefont
  {Duan}}, \bibinfo {author} {\bibfnamefont {H.}~\bibnamefont {Giessen}},\ and\
  \bibinfo {author} {\bibfnamefont {N.}~\bibnamefont {Liu}},\ }\bibfield
  {title} {\bibinfo {title} {Chiral plasmonics},\ }\href
  {https://doi.org/10.1126/sciadv.1602735} {\bibfield  {journal} {\bibinfo
  {journal} {Sci. Adv.}\ }\textbf {\bibinfo {volume} {3}},\ \bibinfo {pages}
  {e1602735} (\bibinfo {year} {2017})}\BibitemShut {NoStop}%
\bibitem [{\citenamefont {Mun}\ \emph {et~al.}(2020)\citenamefont {Mun},
  \citenamefont {Kim}, \citenamefont {Yang}, \citenamefont {Badloe},
  \citenamefont {Ni}, \citenamefont {Chen}, \citenamefont {Qiu},\ and\
  \citenamefont {Rho}}]{ref3}%
  \BibitemOpen
  \bibfield  {author} {\bibinfo {author} {\bibfnamefont {J.}~\bibnamefont
  {Mun}}, \bibinfo {author} {\bibfnamefont {M.}~\bibnamefont {Kim}}, \bibinfo
  {author} {\bibfnamefont {Y.}~\bibnamefont {Yang}}, \bibinfo {author}
  {\bibfnamefont {T.}~\bibnamefont {Badloe}}, \bibinfo {author} {\bibfnamefont
  {J.}~\bibnamefont {Ni}}, \bibinfo {author} {\bibfnamefont {Y.}~\bibnamefont
  {Chen}}, \bibinfo {author} {\bibfnamefont {C.-W.}\ \bibnamefont {Qiu}},\ and\
  \bibinfo {author} {\bibfnamefont {J.}~\bibnamefont {Rho}},\ }\bibfield
  {title} {\bibinfo {title} {Electromagnetic chirality: from fundamentals to
  nontraditional chiroptical phenomena},\ }\href {https://doi.org/10/gmcjhn}
  {\bibfield  {journal} {\bibinfo  {journal} {Light: Sci. Appl.}\ }\textbf
  {\bibinfo {volume} {9}},\ \bibinfo {pages} {139} (\bibinfo {year}
  {2020})}\BibitemShut {NoStop}%
\bibitem [{\citenamefont {Gansel}\ \emph {et~al.}(2009)\citenamefont {Gansel},
  \citenamefont {Thiel}, \citenamefont {Rill}, \citenamefont {Decker},
  \citenamefont {Bade}, \citenamefont {Saile}, \citenamefont {von Freymann},
  \citenamefont {Linden},\ and\ \citenamefont {Wegener}}]{2009gold}%
  \BibitemOpen
  \bibfield  {author} {\bibinfo {author} {\bibfnamefont {J.~K.}\ \bibnamefont
  {Gansel}}, \bibinfo {author} {\bibfnamefont {M.}~\bibnamefont {Thiel}},
  \bibinfo {author} {\bibfnamefont {M.~S.}\ \bibnamefont {Rill}}, \bibinfo
  {author} {\bibfnamefont {M.}~\bibnamefont {Decker}}, \bibinfo {author}
  {\bibfnamefont {K.}~\bibnamefont {Bade}}, \bibinfo {author} {\bibfnamefont
  {V.}~\bibnamefont {Saile}}, \bibinfo {author} {\bibfnamefont
  {G.}~\bibnamefont {von Freymann}}, \bibinfo {author} {\bibfnamefont
  {S.}~\bibnamefont {Linden}},\ and\ \bibinfo {author} {\bibfnamefont
  {M.}~\bibnamefont {Wegener}},\ }\bibfield  {title} {\bibinfo {title} {Gold
  helix photonic metamaterial as broadband circular polarizer},\ }\href
  {https://doi.org/10.1126/science.1177031} {\bibfield  {journal} {\bibinfo
  {journal} {Science}\ }\textbf {\bibinfo {volume} {325}},\ \bibinfo {pages}
  {1513} (\bibinfo {year} {2009})}\BibitemShut {NoStop}%
\bibitem [{\citenamefont {Wu}\ \emph {et~al.}(2011)\citenamefont {Wu},
  \citenamefont {Li}, \citenamefont {Yu}, \citenamefont {Li}, \citenamefont
  {Chen},\ and\ \citenamefont {Chan}}]{wu2011}%
  \BibitemOpen
  \bibfield  {author} {\bibinfo {author} {\bibfnamefont {C.}~\bibnamefont
  {Wu}}, \bibinfo {author} {\bibfnamefont {H.}~\bibnamefont {Li}}, \bibinfo
  {author} {\bibfnamefont {X.}~\bibnamefont {Yu}}, \bibinfo {author}
  {\bibfnamefont {F.}~\bibnamefont {Li}}, \bibinfo {author} {\bibfnamefont
  {H.}~\bibnamefont {Chen}},\ and\ \bibinfo {author} {\bibfnamefont {C.~T.}\
  \bibnamefont {Chan}},\ }\bibfield  {title} {\bibinfo {title} {Metallic helix
  array as a broadband wave plate},\ }\href
  {https://doi.org/https://doi.org/10.1103/PhysRevLett.107.177401} {\bibfield
  {journal} {\bibinfo  {journal} {Phys. Rev. Lett.}\ }\textbf {\bibinfo
  {volume} {107}},\ \bibinfo {pages} {177401} (\bibinfo {year}
  {2011})}\BibitemShut {NoStop}%
\bibitem [{\citenamefont {Wang}\ and\ \citenamefont {Chan}(2014)}]{ref12}%
  \BibitemOpen
  \bibfield  {author} {\bibinfo {author} {\bibfnamefont {S.~B.}\ \bibnamefont
  {Wang}}\ and\ \bibinfo {author} {\bibfnamefont {C.~T.}\ \bibnamefont
  {Chan}},\ }\bibfield  {title} {\bibinfo {title} {Lateral optical force on
  chiral particles near a surface},\ }\href {https://doi.org/10/gkrcg5}
  {\bibfield  {journal} {\bibinfo  {journal} {Nat. Commun.}\ }\textbf {\bibinfo
  {volume} {5}},\ \bibinfo {pages} {3307} (\bibinfo {year} {2014})}\BibitemShut
  {NoStop}%
\bibitem [{\citenamefont {Hayat}\ \emph {et~al.}(2015)\citenamefont {Hayat},
  \citenamefont {Mueller},\ and\ \citenamefont {Capasso}}]{ref13}%
  \BibitemOpen
  \bibfield  {author} {\bibinfo {author} {\bibfnamefont {A.}~\bibnamefont
  {Hayat}}, \bibinfo {author} {\bibfnamefont {J.~P.~B.}\ \bibnamefont
  {Mueller}},\ and\ \bibinfo {author} {\bibfnamefont {F.}~\bibnamefont
  {Capasso}},\ }\bibfield  {title} {\bibinfo {title} {Lateral chirality-sorting
  optical forces},\ }\href {https://doi.org/10/gqms3z} {\bibfield  {journal}
  {\bibinfo  {journal} {Proc. Natl. Acad. Sci. U.S.A.}\ }\textbf {\bibinfo
  {volume} {112}},\ \bibinfo {pages} {13190} (\bibinfo {year}
  {2015})}\BibitemShut {NoStop}%
\bibitem [{\citenamefont {Rahimzadegan}\ \emph {et~al.}(2016)\citenamefont
  {Rahimzadegan}, \citenamefont {Fruhnert}, \citenamefont {Alaee},
  \citenamefont {{Fernandez-Corbaton}},\ and\ \citenamefont
  {Rockstuhl}}]{ref14}%
  \BibitemOpen
  \bibfield  {author} {\bibinfo {author} {\bibfnamefont {A.}~\bibnamefont
  {Rahimzadegan}}, \bibinfo {author} {\bibfnamefont {M.}~\bibnamefont
  {Fruhnert}}, \bibinfo {author} {\bibfnamefont {R.}~\bibnamefont {Alaee}},
  \bibinfo {author} {\bibfnamefont {I.}~\bibnamefont {{Fernandez-Corbaton}}},\
  and\ \bibinfo {author} {\bibfnamefont {C.}~\bibnamefont {Rockstuhl}},\
  }\bibfield  {title} {\bibinfo {title} {Optical force and torque on dipolar
  dual chiral particles},\ }\href {https://doi.org/10/gqms32} {\bibfield
  {journal} {\bibinfo  {journal} {Phys. Rev. B}\ }\textbf {\bibinfo {volume}
  {94}},\ \bibinfo {pages} {125123} (\bibinfo {year} {2016})}\BibitemShut
  {NoStop}%
\bibitem [{\citenamefont {Rechtsman}\ \emph {et~al.}(2013)\citenamefont
  {Rechtsman}, \citenamefont {Zeuner}, \citenamefont {Plotnik}, \citenamefont
  {Lumer}, \citenamefont {Podolsky}, \citenamefont {Dreisow}, \citenamefont
  {Nolte}, \citenamefont {Segev},\ and\ \citenamefont
  {Szameit}}]{2013photonic}%
  \BibitemOpen
  \bibfield  {author} {\bibinfo {author} {\bibfnamefont {M.~C.}\ \bibnamefont
  {Rechtsman}}, \bibinfo {author} {\bibfnamefont {J.~M.}\ \bibnamefont
  {Zeuner}}, \bibinfo {author} {\bibfnamefont {Y.}~\bibnamefont {Plotnik}},
  \bibinfo {author} {\bibfnamefont {Y.}~\bibnamefont {Lumer}}, \bibinfo
  {author} {\bibfnamefont {D.}~\bibnamefont {Podolsky}}, \bibinfo {author}
  {\bibfnamefont {F.}~\bibnamefont {Dreisow}}, \bibinfo {author} {\bibfnamefont
  {S.}~\bibnamefont {Nolte}}, \bibinfo {author} {\bibfnamefont
  {M.}~\bibnamefont {Segev}},\ and\ \bibinfo {author} {\bibfnamefont
  {A.}~\bibnamefont {Szameit}},\ }\bibfield  {title} {\bibinfo {title}
  {Photonic floquet topological insulators},\ }\href
  {https://doi.org/https://doi.org/10.1038/nature12066} {\bibfield  {journal}
  {\bibinfo  {journal} {Nature (London)}\ }\textbf {\bibinfo {volume} {496}},\
  \bibinfo {pages} {196} (\bibinfo {year} {2013})}\BibitemShut {NoStop}%
\bibitem [{\citenamefont {Oh}\ and\ \citenamefont {Hess}(2015)}]{ref1}%
  \BibitemOpen
  \bibfield  {author} {\bibinfo {author} {\bibfnamefont {S.~S.}\ \bibnamefont
  {Oh}}\ and\ \bibinfo {author} {\bibfnamefont {O.}~\bibnamefont {Hess}},\
  }\bibfield  {title} {\bibinfo {title} {Chiral metamaterials: enhancement and
  control of optical activity and circular dichroism},\ }\href
  {https://doi.org/10.1186/s40580-015-0058-2} {\bibfield  {journal} {\bibinfo
  {journal} {Nano Convergence}\ }\textbf {\bibinfo {volume} {2}},\ \bibinfo
  {pages} {24} (\bibinfo {year} {2015})}\BibitemShut {NoStop}%
\bibitem [{\citenamefont {Valev}\ \emph {et~al.}(2013)\citenamefont {Valev},
  \citenamefont {Baumberg}, \citenamefont {Sibilia},\ and\ \citenamefont
  {Verbiest}}]{ref19}%
  \BibitemOpen
  \bibfield  {author} {\bibinfo {author} {\bibfnamefont {V.~K.}\ \bibnamefont
  {Valev}}, \bibinfo {author} {\bibfnamefont {J.~J.}\ \bibnamefont {Baumberg}},
  \bibinfo {author} {\bibfnamefont {C.}~\bibnamefont {Sibilia}},\ and\ \bibinfo
  {author} {\bibfnamefont {T.}~\bibnamefont {Verbiest}},\ }\bibfield  {title}
  {\bibinfo {title} {Chirality and {{Chiroptical Effects}} in {{Plasmonic
  Nanostructures}}: {{Fundamentals}}, {{Recent Progress}}, and {{Outlook}}},\
  }\href {https://doi.org/10/f2b3fx} {\bibfield  {journal} {\bibinfo  {journal}
  {Adv. Mater.}\ }\textbf {\bibinfo {volume} {25}},\ \bibinfo {pages} {2517}
  (\bibinfo {year} {2013})}\BibitemShut {NoStop}%
\bibitem [{\citenamefont {Brullot}\ \emph {et~al.}(2016)\citenamefont
  {Brullot}, \citenamefont {Vanbel}, \citenamefont {Swusten},\ and\
  \citenamefont {Verbiest}}]{ref21}%
  \BibitemOpen
  \bibfield  {author} {\bibinfo {author} {\bibfnamefont {W.}~\bibnamefont
  {Brullot}}, \bibinfo {author} {\bibfnamefont {M.~K.}\ \bibnamefont {Vanbel}},
  \bibinfo {author} {\bibfnamefont {T.}~\bibnamefont {Swusten}},\ and\ \bibinfo
  {author} {\bibfnamefont {T.}~\bibnamefont {Verbiest}},\ }\bibfield  {title}
  {\bibinfo {title} {Resolving enantiomers using the optical angular momentum
  of twisted light},\ }\href {https://doi.org/10/f8hk95} {\bibfield  {journal}
  {\bibinfo  {journal} {Sci. Adv.}\ }\textbf {\bibinfo {volume} {2}},\ \bibinfo
  {pages} {e1501349} (\bibinfo {year} {2016})}\BibitemShut {NoStop}%
\bibitem [{\citenamefont {Forbes}\ and\ \citenamefont {Andrews}(2018)}]{ref22}%
  \BibitemOpen
  \bibfield  {author} {\bibinfo {author} {\bibfnamefont {K.~A.}\ \bibnamefont
  {Forbes}}\ and\ \bibinfo {author} {\bibfnamefont {D.~L.}\ \bibnamefont
  {Andrews}},\ }\bibfield  {title} {\bibinfo {title} {Optical orbital angular
  momentum: twisted light and chirality},\ }\href {https://doi.org/10/gmcjck}
  {\bibfield  {journal} {\bibinfo  {journal} {Opt. Lett.}\ }\textbf {\bibinfo
  {volume} {43}},\ \bibinfo {pages} {435} (\bibinfo {year} {2018})}\BibitemShut
  {NoStop}%
\bibitem [{\citenamefont {Rogacheva}\ \emph {et~al.}(2006)\citenamefont
  {Rogacheva}, \citenamefont {Fedotov}, \citenamefont {Schwanecke},\ and\
  \citenamefont {Zheludev}}]{ref26}%
  \BibitemOpen
  \bibfield  {author} {\bibinfo {author} {\bibfnamefont {A.~V.}\ \bibnamefont
  {Rogacheva}}, \bibinfo {author} {\bibfnamefont {V.~A.}\ \bibnamefont
  {Fedotov}}, \bibinfo {author} {\bibfnamefont {A.~S.}\ \bibnamefont
  {Schwanecke}},\ and\ \bibinfo {author} {\bibfnamefont {N.~I.}\ \bibnamefont
  {Zheludev}},\ }\bibfield  {title} {\bibinfo {title} {Giant {{Gyrotropy}} due
  to {{Electromagnetic-Field Coupling}} in a {{Bilayered Chiral Structure}}},\
  }\href {https://doi.org/10/d59h2b} {\bibfield  {journal} {\bibinfo  {journal}
  {Phys. Rev. Lett.}\ }\textbf {\bibinfo {volume} {97}},\ \bibinfo {pages}
  {177401} (\bibinfo {year} {2006})}\BibitemShut {NoStop}%
\bibitem [{\citenamefont {Cheng}\ \emph {et~al.}(2020)\citenamefont {Cheng},
  \citenamefont {Chen},\ and\ \citenamefont {Luo}}]{ref28}%
  \BibitemOpen
  \bibfield  {author} {\bibinfo {author} {\bibfnamefont {Y.}~\bibnamefont
  {Cheng}}, \bibinfo {author} {\bibfnamefont {F.}~\bibnamefont {Chen}},\ and\
  \bibinfo {author} {\bibfnamefont {H.}~\bibnamefont {Luo}},\ }\bibfield
  {title} {\bibinfo {title} {Multi-band giant circular dichroism based on
  conjugated bilayer twisted-semicircle nanostructure at optical frequency},\
  }\href {https://doi.org/10/gqms3v} {\bibfield  {journal} {\bibinfo  {journal}
  {Phys. Lett. A}\ }\textbf {\bibinfo {volume} {384}},\ \bibinfo {pages}
  {126398} (\bibinfo {year} {2020})}\BibitemShut {NoStop}%
\bibitem [{\citenamefont {Wang}\ \emph {et~al.}(2009)\citenamefont {Wang},
  \citenamefont {Zhou}, \citenamefont {Koschny},\ and\ \citenamefont
  {Soukoulis}}]{ref7}%
  \BibitemOpen
  \bibfield  {author} {\bibinfo {author} {\bibfnamefont {B.}~\bibnamefont
  {Wang}}, \bibinfo {author} {\bibfnamefont {J.}~\bibnamefont {Zhou}}, \bibinfo
  {author} {\bibfnamefont {T.}~\bibnamefont {Koschny}},\ and\ \bibinfo {author}
  {\bibfnamefont {C.~M.}\ \bibnamefont {Soukoulis}},\ }\bibfield  {title}
  {\bibinfo {title} {Nonplanar chiral metamaterials with negative index},\
  }\href {https://doi.org/10/frhnx3} {\bibfield  {journal} {\bibinfo  {journal}
  {Appl. Phys. Lett.}\ }\textbf {\bibinfo {volume} {94}},\ \bibinfo {pages}
  {151112} (\bibinfo {year} {2009})}\BibitemShut {NoStop}%
\bibitem [{\citenamefont {Saba}\ \emph {et~al.}(2011)\citenamefont {Saba},
  \citenamefont {Thiel}, \citenamefont {Turner}, \citenamefont {Hyde},
  \citenamefont {Gu}, \citenamefont {{Grosse-Brauckmann}}, \citenamefont
  {Neshev}, \citenamefont {Mecke},\ and\ \citenamefont
  {{Schröder-Turk}}}]{ref29}%
  \BibitemOpen
  \bibfield  {author} {\bibinfo {author} {\bibfnamefont {M.}~\bibnamefont
  {Saba}}, \bibinfo {author} {\bibfnamefont {M.}~\bibnamefont {Thiel}},
  \bibinfo {author} {\bibfnamefont {M.~D.}\ \bibnamefont {Turner}}, \bibinfo
  {author} {\bibfnamefont {S.~T.}\ \bibnamefont {Hyde}}, \bibinfo {author}
  {\bibfnamefont {M.}~\bibnamefont {Gu}}, \bibinfo {author} {\bibfnamefont
  {K.}~\bibnamefont {{Grosse-Brauckmann}}}, \bibinfo {author} {\bibfnamefont
  {D.~N.}\ \bibnamefont {Neshev}}, \bibinfo {author} {\bibfnamefont
  {K.}~\bibnamefont {Mecke}},\ and\ \bibinfo {author} {\bibfnamefont {G.~E.}\
  \bibnamefont {{Schröder-Turk}}},\ }\bibfield  {title} {\bibinfo {title}
  {Circular {{Dichroism}} in {{Biological Photonic Crystals}} and {{Cubic
  Chiral Nets}}},\ }\href {https://doi.org/10/fk3s8v} {\bibfield  {journal}
  {\bibinfo  {journal} {Phys. Rev. Lett.}\ }\textbf {\bibinfo {volume} {106}},\
  \bibinfo {pages} {103902} (\bibinfo {year} {2011})}\BibitemShut {NoStop}%
\bibitem [{\citenamefont {Kilchoer}\ \emph {et~al.}(2020)\citenamefont
  {Kilchoer}, \citenamefont {Abdollahi}, \citenamefont {Dolan}, \citenamefont
  {Abdelrahman}, \citenamefont {Saba}, \citenamefont {Wiesner}, \citenamefont
  {Steiner}, \citenamefont {Gunkel},\ and\ \citenamefont {Wilts}}]{ref31}%
  \BibitemOpen
  \bibfield  {author} {\bibinfo {author} {\bibfnamefont {C.}~\bibnamefont
  {Kilchoer}}, \bibinfo {author} {\bibfnamefont {N.}~\bibnamefont {Abdollahi}},
  \bibinfo {author} {\bibfnamefont {J.~A.}\ \bibnamefont {Dolan}}, \bibinfo
  {author} {\bibfnamefont {D.}~\bibnamefont {Abdelrahman}}, \bibinfo {author}
  {\bibfnamefont {M.}~\bibnamefont {Saba}}, \bibinfo {author} {\bibfnamefont
  {U.}~\bibnamefont {Wiesner}}, \bibinfo {author} {\bibfnamefont
  {U.}~\bibnamefont {Steiner}}, \bibinfo {author} {\bibfnamefont
  {I.}~\bibnamefont {Gunkel}},\ and\ \bibinfo {author} {\bibfnamefont {B.~D.}\
  \bibnamefont {Wilts}},\ }\bibfield  {title} {\bibinfo {title} {Strong
  {{Circular Dichroism}} in {{Single Gyroid Optical Metamaterials}}},\ }\href
  {https://doi.org/10/gm8pgd} {\bibfield  {journal} {\bibinfo  {journal} {Adv.
  Opt. Mater.}\ }\textbf {\bibinfo {volume} {8}},\ \bibinfo {pages} {1902131}
  (\bibinfo {year} {2020})}\BibitemShut {NoStop}%
\bibitem [{\citenamefont {Khanikaev}\ \emph {et~al.}(2016)\citenamefont
  {Khanikaev}, \citenamefont {Arju}, \citenamefont {Fan}, \citenamefont
  {Purtseladze}, \citenamefont {Lu}, \citenamefont {Lee}, \citenamefont
  {Sarriugarte}, \citenamefont {Schnell}, \citenamefont {Hillenbrand},
  \citenamefont {Belkin},\ and\ \citenamefont {Shvets}}]{ref32}%
  \BibitemOpen
  \bibfield  {author} {\bibinfo {author} {\bibfnamefont {A.~B.}\ \bibnamefont
  {Khanikaev}}, \bibinfo {author} {\bibfnamefont {N.}~\bibnamefont {Arju}},
  \bibinfo {author} {\bibfnamefont {Z.}~\bibnamefont {Fan}}, \bibinfo {author}
  {\bibfnamefont {D.}~\bibnamefont {Purtseladze}}, \bibinfo {author}
  {\bibfnamefont {F.}~\bibnamefont {Lu}}, \bibinfo {author} {\bibfnamefont
  {J.}~\bibnamefont {Lee}}, \bibinfo {author} {\bibfnamefont {P.}~\bibnamefont
  {Sarriugarte}}, \bibinfo {author} {\bibfnamefont {M.}~\bibnamefont
  {Schnell}}, \bibinfo {author} {\bibfnamefont {R.}~\bibnamefont
  {Hillenbrand}}, \bibinfo {author} {\bibfnamefont {M.~A.}\ \bibnamefont
  {Belkin}},\ and\ \bibinfo {author} {\bibfnamefont {G.}~\bibnamefont
  {Shvets}},\ }\bibfield  {title} {\bibinfo {title} {Experimental demonstration
  of the microscopic origin of circular dichroism in two-dimensional
  metamaterials},\ }\href {https://doi.org/10/f8tqhb} {\bibfield  {journal}
  {\bibinfo  {journal} {Nat. Commun.}\ }\textbf {\bibinfo {volume} {7}},\
  \bibinfo {pages} {12045} (\bibinfo {year} {2016})}\BibitemShut {NoStop}%
\bibitem [{\citenamefont {Shi}\ \emph {et~al.}(2022)\citenamefont {Shi},
  \citenamefont {Deng}, \citenamefont {Geng}, \citenamefont {Zeng},
  \citenamefont {Zeng}, \citenamefont {Hu}, \citenamefont {Overvig},
  \citenamefont {Li}, \citenamefont {Qiu}, \citenamefont {Al{\`u}},
  \citenamefont {Kivshar},\ and\ \citenamefont {Li}}]{shi2022}%
  \BibitemOpen
  \bibfield  {author} {\bibinfo {author} {\bibfnamefont {T.}~\bibnamefont
  {Shi}}, \bibinfo {author} {\bibfnamefont {Z.-L.}\ \bibnamefont {Deng}},
  \bibinfo {author} {\bibfnamefont {G.}~\bibnamefont {Geng}}, \bibinfo {author}
  {\bibfnamefont {X.}~\bibnamefont {Zeng}}, \bibinfo {author} {\bibfnamefont
  {Y.}~\bibnamefont {Zeng}}, \bibinfo {author} {\bibfnamefont {G.}~\bibnamefont
  {Hu}}, \bibinfo {author} {\bibfnamefont {A.}~\bibnamefont {Overvig}},
  \bibinfo {author} {\bibfnamefont {J.}~\bibnamefont {Li}}, \bibinfo {author}
  {\bibfnamefont {C.-W.}\ \bibnamefont {Qiu}}, \bibinfo {author} {\bibfnamefont
  {A.}~\bibnamefont {Al{\`u}}}, \bibinfo {author} {\bibfnamefont {Y.~S.}\
  \bibnamefont {Kivshar}},\ and\ \bibinfo {author} {\bibfnamefont
  {X.}~\bibnamefont {Li}},\ }\bibfield  {title} {\bibinfo {title} {Planar
  chiral metasurfaces with maximal and tunable chiroptical response driven by
  bound states in the continuum},\ }\href
  {https://doi.org/https://doi.org/10.1038/s41467-022-31877-1} {\bibfield
  {journal} {\bibinfo  {journal} {Nat. Commun.}\ }\textbf {\bibinfo {volume}
  {13}},\ \bibinfo {pages} {4111} (\bibinfo {year} {2022})}\BibitemShut
  {NoStop}%
\bibitem [{\citenamefont {Berova}\ \emph {et~al.}(2000)\citenamefont {Berova},
  \citenamefont {Nakanishi},\ and\ \citenamefont {Woody}}]{ref34}%
  \BibitemOpen
  \bibfield  {author} {\bibinfo {author} {\bibfnamefont {N.}~\bibnamefont
  {Berova}}, \bibinfo {author} {\bibfnamefont {K.}~\bibnamefont {Nakanishi}},\
  and\ \bibinfo {author} {\bibfnamefont {R.~W.}\ \bibnamefont {Woody}},\
  }\href@noop {} {\emph {\bibinfo {title} {Circular {{Dichroism}}:
  {{Principles}} and {{Applications}}}}}\ (\bibinfo  {publisher} {{Wiley, New
  York}},\ \bibinfo {year} {2000})\BibitemShut {NoStop}%
\bibitem [{\citenamefont {Micsonai}\ \emph {et~al.}(2015)\citenamefont
  {Micsonai}, \citenamefont {Wien}, \citenamefont {Kernya}, \citenamefont
  {Lee}, \citenamefont {Goto}, \citenamefont {Réfrégiers},\ and\
  \citenamefont {Kardos}}]{ref37}%
  \BibitemOpen
  \bibfield  {author} {\bibinfo {author} {\bibfnamefont {A.}~\bibnamefont
  {Micsonai}}, \bibinfo {author} {\bibfnamefont {F.}~\bibnamefont {Wien}},
  \bibinfo {author} {\bibfnamefont {L.}~\bibnamefont {Kernya}}, \bibinfo
  {author} {\bibfnamefont {Y.-H.}\ \bibnamefont {Lee}}, \bibinfo {author}
  {\bibfnamefont {Y.}~\bibnamefont {Goto}}, \bibinfo {author} {\bibfnamefont
  {M.}~\bibnamefont {Réfrégiers}},\ and\ \bibinfo {author} {\bibfnamefont
  {J.}~\bibnamefont {Kardos}},\ }\bibfield  {title} {\bibinfo {title} {Accurate
  secondary structure prediction and fold recognition for circular dichroism
  spectroscopy},\ }\href {https://doi.org/10/f7gcz5} {\bibfield  {journal}
  {\bibinfo  {journal} {Proc. Natl. Acad. Sci. U.S.A.}\ }\textbf {\bibinfo
  {volume} {112}},\ \bibinfo {pages} {E3095} (\bibinfo {year}
  {2015})}\BibitemShut {NoStop}%
\bibitem [{\citenamefont {Vinegrad}\ \emph {et~al.}(2018)\citenamefont
  {Vinegrad}, \citenamefont {Vestler}, \citenamefont {Ben-Moshe}, \citenamefont
  {Barnea}, \citenamefont {Markovich},\ and\ \citenamefont
  {Cheshnovsky}}]{2018circular}%
  \BibitemOpen
  \bibfield  {author} {\bibinfo {author} {\bibfnamefont {E.}~\bibnamefont
  {Vinegrad}}, \bibinfo {author} {\bibfnamefont {D.}~\bibnamefont {Vestler}},
  \bibinfo {author} {\bibfnamefont {A.}~\bibnamefont {Ben-Moshe}}, \bibinfo
  {author} {\bibfnamefont {A.~R.}\ \bibnamefont {Barnea}}, \bibinfo {author}
  {\bibfnamefont {G.}~\bibnamefont {Markovich}},\ and\ \bibinfo {author}
  {\bibfnamefont {O.}~\bibnamefont {Cheshnovsky}},\ }\bibfield  {title}
  {\bibinfo {title} {Circular dichroism of single particles},\ }\href
  {https://doi.org/https://doi.org/10.1021/acsphotonics.8b00016} {\bibfield
  {journal} {\bibinfo  {journal} {ACS Photonics}\ }\textbf {\bibinfo {volume}
  {5}},\ \bibinfo {pages} {2151} (\bibinfo {year} {2018})}\BibitemShut
  {NoStop}%
\bibitem [{\citenamefont {Jia}\ \emph {et~al.}(2022)\citenamefont {Jia},
  \citenamefont {Peng}, \citenamefont {Cheng},\ and\ \citenamefont
  {Wang}}]{2022chiral}%
  \BibitemOpen
  \bibfield  {author} {\bibinfo {author} {\bibfnamefont {S.}~\bibnamefont
  {Jia}}, \bibinfo {author} {\bibfnamefont {J.}~\bibnamefont {Peng}}, \bibinfo
  {author} {\bibfnamefont {Y.}~\bibnamefont {Cheng}},\ and\ \bibinfo {author}
  {\bibfnamefont {S.}~\bibnamefont {Wang}},\ }\bibfield  {title} {\bibinfo
  {title} {Chiral discrimination by polarization singularities of a metal
  sphere},\ }\href
  {https://doi.org/https://doi.org/10.1103/PhysRevA.105.033513} {\bibfield
  {journal} {\bibinfo  {journal} {Phys. Rev. A}\ }\textbf {\bibinfo {volume}
  {105}},\ \bibinfo {pages} {033513} (\bibinfo {year} {2022})}\BibitemShut
  {NoStop}%
\bibitem [{\citenamefont {Jiang}\ \emph {et~al.}(2016)\citenamefont {Jiang},
  \citenamefont {Li}, \citenamefont {Liang}, \citenamefont {Cheng},\ and\
  \citenamefont {Zhang}}]{ref40}%
  \BibitemOpen
  \bibfield  {author} {\bibinfo {author} {\bibfnamefont {X.}~\bibnamefont
  {Jiang}}, \bibinfo {author} {\bibfnamefont {Y.}~\bibnamefont {Li}}, \bibinfo
  {author} {\bibfnamefont {B.}~\bibnamefont {Liang}}, \bibinfo {author}
  {\bibfnamefont {J.-c.}\ \bibnamefont {Cheng}},\ and\ \bibinfo {author}
  {\bibfnamefont {L.}~\bibnamefont {Zhang}},\ }\bibfield  {title} {\bibinfo
  {title} {Convert {{Acoustic Resonances}} to {{Orbital Angular Momentum}}},\
  }\href {https://doi.org/10/gp9qs3} {\bibfield  {journal} {\bibinfo  {journal}
  {Phys. Rev. Lett.}\ }\textbf {\bibinfo {volume} {117}},\ \bibinfo {pages}
  {034301} (\bibinfo {year} {2016})}\BibitemShut {NoStop}%
\bibitem [{\citenamefont {Esfahlani}\ \emph {et~al.}(2017)\citenamefont
  {Esfahlani}, \citenamefont {Lissek},\ and\ \citenamefont {Mosig}}]{ref41}%
  \BibitemOpen
  \bibfield  {author} {\bibinfo {author} {\bibfnamefont {H.}~\bibnamefont
  {Esfahlani}}, \bibinfo {author} {\bibfnamefont {H.}~\bibnamefont {Lissek}},\
  and\ \bibinfo {author} {\bibfnamefont {J.~R.}\ \bibnamefont {Mosig}},\
  }\bibfield  {title} {\bibinfo {title} {Generation of acoustic helical
  wavefronts using metasurfaces},\ }\href {https://doi.org/10/gg9fht}
  {\bibfield  {journal} {\bibinfo  {journal} {Phys. Rev. B}\ }\textbf {\bibinfo
  {volume} {95}},\ \bibinfo {pages} {024312} (\bibinfo {year}
  {2017})}\BibitemShut {NoStop}%
\bibitem [{\citenamefont {Zhou}\ \emph {et~al.}(2019)\citenamefont {Zhou},
  \citenamefont {Li}, \citenamefont {Guo},\ and\ \citenamefont {Guo}}]{ref42}%
  \BibitemOpen
  \bibfield  {author} {\bibinfo {author} {\bibfnamefont {H.}~\bibnamefont
  {Zhou}}, \bibinfo {author} {\bibfnamefont {J.}~\bibnamefont {Li}}, \bibinfo
  {author} {\bibfnamefont {K.}~\bibnamefont {Guo}},\ and\ \bibinfo {author}
  {\bibfnamefont {Z.}~\bibnamefont {Guo}},\ }\bibfield  {title} {\bibinfo
  {title} {Generation of acoustic vortex beams with designed fermat's spiral
  diffraction grating},\ }\href {https://doi.org/10/gjzm77} {\bibfield
  {journal} {\bibinfo  {journal} {J. Acoust. Soc. Am.}\ }\textbf {\bibinfo
  {volume} {146}},\ \bibinfo {pages} {4237} (\bibinfo {year}
  {2019})}\BibitemShut {NoStop}%
\bibitem [{\citenamefont {Wang}\ \emph {et~al.}(2018)\citenamefont {Wang},
  \citenamefont {Ma},\ and\ \citenamefont {Chan}}]{ref46}%
  \BibitemOpen
  \bibfield  {author} {\bibinfo {author} {\bibfnamefont {S.}~\bibnamefont
  {Wang}}, \bibinfo {author} {\bibfnamefont {G.}~\bibnamefont {Ma}},\ and\
  \bibinfo {author} {\bibfnamefont {C.~T.}\ \bibnamefont {Chan}},\ }\bibfield
  {title} {\bibinfo {title} {Topological transport of sound mediated by
  spin-redirection geometric phase},\ }\href {https://doi.org/10/gc2498}
  {\bibfield  {journal} {\bibinfo  {journal} {Sci. Adv.}\ }\textbf {\bibinfo
  {volume} {4}},\ \bibinfo {pages} {eaaq1475} (\bibinfo {year}
  {2018})}\BibitemShut {NoStop}%
\bibitem [{\citenamefont {Wang}\ \emph
  {et~al.}(2021{\natexlab{a}})\citenamefont {Wang}, \citenamefont {Tan},
  \citenamefont {Liang}, \citenamefont {Ma}, \citenamefont {Wang},\ and\
  \citenamefont {Cheng}}]{ref54}%
  \BibitemOpen
  \bibfield  {author} {\bibinfo {author} {\bibfnamefont {W.}~\bibnamefont
  {Wang}}, \bibinfo {author} {\bibfnamefont {Y.}~\bibnamefont {Tan}}, \bibinfo
  {author} {\bibfnamefont {B.}~\bibnamefont {Liang}}, \bibinfo {author}
  {\bibfnamefont {G.}~\bibnamefont {Ma}}, \bibinfo {author} {\bibfnamefont
  {S.}~\bibnamefont {Wang}},\ and\ \bibinfo {author} {\bibfnamefont
  {J.}~\bibnamefont {Cheng}},\ }\bibfield  {title} {\bibinfo {title}
  {Generalized momentum conservation and {{Fedorov-Imbert}} linear shift of
  acoustic vortex beams at a metasurface},\ }\href {https://doi.org/10/gngfvh}
  {\bibfield  {journal} {\bibinfo  {journal} {Phys. Rev. B}\ }\textbf {\bibinfo
  {volume} {104}},\ \bibinfo {pages} {174301} (\bibinfo {year}
  {2021}{\natexlab{a}})}\BibitemShut {NoStop}%
\bibitem [{\citenamefont {Fan}\ and\ \citenamefont {Zhang}(2021)}]{ref55}%
  \BibitemOpen
  \bibfield  {author} {\bibinfo {author} {\bibfnamefont {X.-D.}\ \bibnamefont
  {Fan}}\ and\ \bibinfo {author} {\bibfnamefont {L.}~\bibnamefont {Zhang}},\
  }\bibfield  {title} {\bibinfo {title} {Acoustic orbital angular momentum
  {{Hall}} effect and realization using a metasurface},\ }\href
  {https://doi.org/10/gjmfbs} {\bibfield  {journal} {\bibinfo  {journal} {Phys.
  Rev. Research}\ }\textbf {\bibinfo {volume} {3}},\ \bibinfo {pages} {013251}
  (\bibinfo {year} {2021})}\BibitemShut {NoStop}%
\bibitem [{\citenamefont {Fu}\ \emph {et~al.}(2020)\citenamefont {Fu},
  \citenamefont {Shen}, \citenamefont {Zhu}, \citenamefont {Li}, \citenamefont
  {Liu}, \citenamefont {Cummer},\ and\ \citenamefont {Xu}}]{ref53}%
  \BibitemOpen
  \bibfield  {author} {\bibinfo {author} {\bibfnamefont {Y.}~\bibnamefont
  {Fu}}, \bibinfo {author} {\bibfnamefont {C.}~\bibnamefont {Shen}}, \bibinfo
  {author} {\bibfnamefont {X.}~\bibnamefont {Zhu}}, \bibinfo {author}
  {\bibfnamefont {J.}~\bibnamefont {Li}}, \bibinfo {author} {\bibfnamefont
  {Y.}~\bibnamefont {Liu}}, \bibinfo {author} {\bibfnamefont {S.~A.}\
  \bibnamefont {Cummer}},\ and\ \bibinfo {author} {\bibfnamefont
  {Y.}~\bibnamefont {Xu}},\ }\bibfield  {title} {\bibinfo {title} {Sound vortex
  diffraction via topological charge in phase gradient metagratings},\ }\href
  {https://doi.org/10/gnzm44} {\bibfield  {journal} {\bibinfo  {journal} {Sci.
  Adv.}\ }\textbf {\bibinfo {volume} {6}},\ \bibinfo {pages} {eaba9876}
  (\bibinfo {year} {2020})}\BibitemShut {NoStop}%
\bibitem [{\citenamefont {Zou}\ \emph {et~al.}(2020)\citenamefont {Zou},
  \citenamefont {Lirette},\ and\ \citenamefont {Zhang}}]{ref56}%
  \BibitemOpen
  \bibfield  {author} {\bibinfo {author} {\bibfnamefont {Z.}~\bibnamefont
  {Zou}}, \bibinfo {author} {\bibfnamefont {R.}~\bibnamefont {Lirette}},\ and\
  \bibinfo {author} {\bibfnamefont {L.}~\bibnamefont {Zhang}},\ }\bibfield
  {title} {\bibinfo {title} {Orbital {{Angular Momentum Reversal}} and
  {{Asymmetry}} in {{Acoustic Vortex Beam Reflection}}},\ }\href
  {https://doi.org/10.1103/physrevlett.125.074301} {\bibfield  {journal}
  {\bibinfo  {journal} {Phys. Rev. Lett.}\ }\textbf {\bibinfo {volume} {125}},\
  \bibinfo {pages} {074301} (\bibinfo {year} {2020})}\BibitemShut {NoStop}%
\bibitem [{\citenamefont {Tong}\ and\ \citenamefont {Wang}(2022)}]{ref45}%
  \BibitemOpen
  \bibfield  {author} {\bibinfo {author} {\bibfnamefont {Q.}~\bibnamefont
  {Tong}}\ and\ \bibinfo {author} {\bibfnamefont {S.}~\bibnamefont {Wang}},\
  }\bibfield  {title} {\bibinfo {title} {Acoustic helical dichroism in a
  one-dimensional lattice of chiral resonators},\ }\href
  {https://doi.org/10/gqms3j} {\bibfield  {journal} {\bibinfo  {journal} {Phys.
  Rev. B}\ }\textbf {\bibinfo {volume} {105}},\ \bibinfo {pages} {024111}
  (\bibinfo {year} {2022})}\BibitemShut {NoStop}%
\bibitem [{\citenamefont {Hong}\ \emph {et~al.}(2017)\citenamefont {Hong},
  \citenamefont {Yin}, \citenamefont {Zhai}, \citenamefont {Yan}, \citenamefont
  {Wang}, \citenamefont {Zhang},\ and\ \citenamefont {Drinkwater}}]{ref47}%
  \BibitemOpen
  \bibfield  {author} {\bibinfo {author} {\bibfnamefont {Z.~Y.}\ \bibnamefont
  {Hong}}, \bibinfo {author} {\bibfnamefont {J.~F.}\ \bibnamefont {Yin}},
  \bibinfo {author} {\bibfnamefont {W.}~\bibnamefont {Zhai}}, \bibinfo {author}
  {\bibfnamefont {N.}~\bibnamefont {Yan}}, \bibinfo {author} {\bibfnamefont
  {W.~L.}\ \bibnamefont {Wang}}, \bibinfo {author} {\bibfnamefont
  {J.}~\bibnamefont {Zhang}},\ and\ \bibinfo {author} {\bibfnamefont {B.~W.}\
  \bibnamefont {Drinkwater}},\ }\bibfield  {title} {\bibinfo {title} {Dynamics
  of levitated objects in acoustic vortex fields},\ }\href
  {https://doi.org/10/gbr3d8} {\bibfield  {journal} {\bibinfo  {journal} {Sci.
  Rep.}\ }\textbf {\bibinfo {volume} {7}},\ \bibinfo {pages} {7093} (\bibinfo
  {year} {2017})}\BibitemShut {NoStop}%
\bibitem [{\citenamefont {Hong}\ \emph {et~al.}(2020)\citenamefont {Hong},
  \citenamefont {Yin}, \citenamefont {Zhang},\ and\ \citenamefont
  {Yan}}]{ref48}%
  \BibitemOpen
  \bibfield  {author} {\bibinfo {author} {\bibfnamefont {Z.~Y.}\ \bibnamefont
  {Hong}}, \bibinfo {author} {\bibfnamefont {J.~F.}\ \bibnamefont {Yin}},
  \bibinfo {author} {\bibfnamefont {B.~W.}\ \bibnamefont {Zhang}},\ and\
  \bibinfo {author} {\bibfnamefont {N.}~\bibnamefont {Yan}},\ }\bibfield
  {title} {\bibinfo {title} {Vortex-field acoustic levitation in tubes},\
  }\href {https://doi.org/10/gqm2jh} {\bibfield  {journal} {\bibinfo  {journal}
  {J. Appl. Phys.}\ }\textbf {\bibinfo {volume} {128}},\ \bibinfo {pages}
  {104901} (\bibinfo {year} {2020})}\BibitemShut {NoStop}%
\bibitem [{\citenamefont {Marzo}\ \emph {et~al.}(2018)\citenamefont {Marzo},
  \citenamefont {Caleap},\ and\ \citenamefont {Drinkwater}}]{ref44}%
  \BibitemOpen
  \bibfield  {author} {\bibinfo {author} {\bibfnamefont {A.}~\bibnamefont
  {Marzo}}, \bibinfo {author} {\bibfnamefont {M.}~\bibnamefont {Caleap}},\ and\
  \bibinfo {author} {\bibfnamefont {B.~W.}\ \bibnamefont {Drinkwater}},\
  }\bibfield  {title} {\bibinfo {title} {Acoustic {{Virtual Vortices}} with
  {{Tunable Orbital Angular Momentum}} for {{Trapping}} of {{Mie Particles}}},\
  }\href {https://doi.org/10.1103/physrevlett.120.044301} {\bibfield  {journal}
  {\bibinfo  {journal} {Phys. Rev. Lett.}\ }\textbf {\bibinfo {volume} {120}},\
  \bibinfo {pages} {044301} (\bibinfo {year} {2018})}\BibitemShut {NoStop}%
\bibitem [{\citenamefont {Baudoin}\ \emph {et~al.}(2019)\citenamefont
  {Baudoin}, \citenamefont {Gerbedoen}, \citenamefont {Riaud}, \citenamefont
  {Matar}, \citenamefont {Smagin},\ and\ \citenamefont {Thomas}}]{ref49}%
  \BibitemOpen
  \bibfield  {author} {\bibinfo {author} {\bibfnamefont {M.}~\bibnamefont
  {Baudoin}}, \bibinfo {author} {\bibfnamefont {J.-C.}\ \bibnamefont
  {Gerbedoen}}, \bibinfo {author} {\bibfnamefont {A.}~\bibnamefont {Riaud}},
  \bibinfo {author} {\bibfnamefont {O.~B.}\ \bibnamefont {Matar}}, \bibinfo
  {author} {\bibfnamefont {N.}~\bibnamefont {Smagin}},\ and\ \bibinfo {author}
  {\bibfnamefont {J.-L.}\ \bibnamefont {Thomas}},\ }\bibfield  {title}
  {\bibinfo {title} {Folding a focalized acoustical vortex on a flat
  holographic transducer: Miniaturized selective acoustical tweezers},\ }\href
  {https://doi.org/10/gj325n} {\bibfield  {journal} {\bibinfo  {journal} {Sci.
  Adv.}\ }\textbf {\bibinfo {volume} {5}},\ \bibinfo {pages} {eaav1967}
  (\bibinfo {year} {2019})}\BibitemShut {NoStop}%
\bibitem [{\citenamefont {Lo}\ \emph {et~al.}(2021)\citenamefont {Lo},
  \citenamefont {Fan}, \citenamefont {Ho}, \citenamefont {Lin},\ and\
  \citenamefont {Yeh}}]{ref50}%
  \BibitemOpen
  \bibfield  {author} {\bibinfo {author} {\bibfnamefont {W.-C.}\ \bibnamefont
  {Lo}}, \bibinfo {author} {\bibfnamefont {C.-H.}\ \bibnamefont {Fan}},
  \bibinfo {author} {\bibfnamefont {Y.-J.}\ \bibnamefont {Ho}}, \bibinfo
  {author} {\bibfnamefont {C.-W.}\ \bibnamefont {Lin}},\ and\ \bibinfo {author}
  {\bibfnamefont {C.-K.}\ \bibnamefont {Yeh}},\ }\bibfield  {title} {\bibinfo
  {title} {Tornado-inspired acoustic vortex tweezer for trapping and
  manipulating microbubbles},\ }\href {https://doi.org/10/gqm2jp} {\bibfield
  {journal} {\bibinfo  {journal} {Proc. Natl. Acad. Sci. U.S.A.}\ }\textbf
  {\bibinfo {volume} {118}},\ \bibinfo {pages} {e2023188118} (\bibinfo {year}
  {2021})}\BibitemShut {NoStop}%
\bibitem [{\citenamefont {Li}\ \emph {et~al.}(2018)\citenamefont {Li},
  \citenamefont {Guo}, \citenamefont {Tu}, \citenamefont {Ma}, \citenamefont
  {Guo}, \citenamefont {Zhang},\ and\ \citenamefont {Sapozhnikov}}]{ref51}%
  \BibitemOpen
  \bibfield  {author} {\bibinfo {author} {\bibfnamefont {Y.}~\bibnamefont
  {Li}}, \bibinfo {author} {\bibfnamefont {G.}~\bibnamefont {Guo}}, \bibinfo
  {author} {\bibfnamefont {J.}~\bibnamefont {Tu}}, \bibinfo {author}
  {\bibfnamefont {Q.}~\bibnamefont {Ma}}, \bibinfo {author} {\bibfnamefont
  {X.}~\bibnamefont {Guo}}, \bibinfo {author} {\bibfnamefont {D.}~\bibnamefont
  {Zhang}},\ and\ \bibinfo {author} {\bibfnamefont {O.~A.}\ \bibnamefont
  {Sapozhnikov}},\ }\bibfield  {title} {\bibinfo {title} {Acoustic radiation
  torque of an acoustic-vortex spanner exerted on axisymmetric objects},\
  }\href {https://doi.org/10/gqm9z2} {\bibfield  {journal} {\bibinfo  {journal}
  {Appl. Phys. Lett.}\ }\textbf {\bibinfo {volume} {112}},\ \bibinfo {pages}
  {254101} (\bibinfo {year} {2018})}\BibitemShut {NoStop}%
\bibitem [{\citenamefont {Wunenburger}\ \emph {et~al.}(2015)\citenamefont
  {Wunenburger}, \citenamefont {Lozano},\ and\ \citenamefont
  {Brasselet}}]{ref52}%
  \BibitemOpen
  \bibfield  {author} {\bibinfo {author} {\bibfnamefont {R.}~\bibnamefont
  {Wunenburger}}, \bibinfo {author} {\bibfnamefont {J.~I.~V.}\ \bibnamefont
  {Lozano}},\ and\ \bibinfo {author} {\bibfnamefont {E.}~\bibnamefont
  {Brasselet}},\ }\bibfield  {title} {\bibinfo {title} {Acoustic orbital
  angular momentum transfer to matter by chiral scattering},\ }\href
  {https://doi.org/https://doi.org/10.1088/1367-2630/17/10/103022} {\bibfield
  {journal} {\bibinfo  {journal} {New J. Phys.}\ }\textbf {\bibinfo {volume}
  {17}},\ \bibinfo {pages} {103022} (\bibinfo {year} {2015})}\BibitemShut
  {NoStop}%
\bibitem [{\citenamefont {Bliokh}\ and\ \citenamefont {Nori}(2019)}]{ref58}%
  \BibitemOpen
  \bibfield  {author} {\bibinfo {author} {\bibfnamefont {K.~Y.}\ \bibnamefont
  {Bliokh}}\ and\ \bibinfo {author} {\bibfnamefont {F.}~\bibnamefont {Nori}},\
  }\bibfield  {title} {\bibinfo {title} {Spin and orbital angular momenta of
  acoustic beams},\ }\href {https://doi.org/10/gf9rt6} {\bibfield  {journal}
  {\bibinfo  {journal} {Phys. Rev. B}\ }\textbf {\bibinfo {volume} {99}},\
  \bibinfo {pages} {174310} (\bibinfo {year} {2019})}\BibitemShut {NoStop}%
\bibitem [{\citenamefont {Long}\ \emph {et~al.}(2020)\citenamefont {Long},
  \citenamefont {Zhang}, \citenamefont {Yang}, \citenamefont {Ge},
  \citenamefont {Chen},\ and\ \citenamefont {Ren}}]{ref60}%
  \BibitemOpen
  \bibfield  {author} {\bibinfo {author} {\bibfnamefont {Y.}~\bibnamefont
  {Long}}, \bibinfo {author} {\bibfnamefont {D.}~\bibnamefont {Zhang}},
  \bibinfo {author} {\bibfnamefont {C.}~\bibnamefont {Yang}}, \bibinfo {author}
  {\bibfnamefont {J.}~\bibnamefont {Ge}}, \bibinfo {author} {\bibfnamefont
  {H.}~\bibnamefont {Chen}},\ and\ \bibinfo {author} {\bibfnamefont
  {J.}~\bibnamefont {Ren}},\ }\bibfield  {title} {\bibinfo {title} {Realization
  of acoustic spin transport in metasurface waveguides},\ }\href
  {https://doi.org/10/gmjppn} {\bibfield  {journal} {\bibinfo  {journal} {Nat.
  Commun.}\ }\textbf {\bibinfo {volume} {11}},\ \bibinfo {pages} {4716}
  (\bibinfo {year} {2020})}\BibitemShut {NoStop}%
\bibitem [{\citenamefont {Wang}\ \emph
  {et~al.}(2021{\natexlab{b}})\citenamefont {Wang}, \citenamefont {Zhang},
  \citenamefont {Wang}, \citenamefont {Tong}, \citenamefont {Li},\ and\
  \citenamefont {Ma}}]{ref61}%
  \BibitemOpen
  \bibfield  {author} {\bibinfo {author} {\bibfnamefont {S.}~\bibnamefont
  {Wang}}, \bibinfo {author} {\bibfnamefont {G.}~\bibnamefont {Zhang}},
  \bibinfo {author} {\bibfnamefont {X.}~\bibnamefont {Wang}}, \bibinfo {author}
  {\bibfnamefont {Q.}~\bibnamefont {Tong}}, \bibinfo {author} {\bibfnamefont
  {J.}~\bibnamefont {Li}},\ and\ \bibinfo {author} {\bibfnamefont
  {G.}~\bibnamefont {Ma}},\ }\bibfield  {title} {\bibinfo {title} {Spin-orbit
  interactions of transverse sound},\ }\href {https://doi.org/10/gpcjv9}
  {\bibfield  {journal} {\bibinfo  {journal} {Nat. Commun.}\ }\textbf {\bibinfo
  {volume} {12}},\ \bibinfo {pages} {6125} (\bibinfo {year}
  {2021}{\natexlab{b}})}\BibitemShut {NoStop}%
\bibitem [{\citenamefont {Decker}\ \emph {et~al.}(2010)\citenamefont {Decker},
  \citenamefont {Zhao}, \citenamefont {Soukoulis}, \citenamefont {Linden},\
  and\ \citenamefont {Wegener}}]{2010twisted}%
  \BibitemOpen
  \bibfield  {author} {\bibinfo {author} {\bibfnamefont {M.}~\bibnamefont
  {Decker}}, \bibinfo {author} {\bibfnamefont {R.}~\bibnamefont {Zhao}},
  \bibinfo {author} {\bibfnamefont {C.~M.}\ \bibnamefont {Soukoulis}}, \bibinfo
  {author} {\bibfnamefont {S.}~\bibnamefont {Linden}},\ and\ \bibinfo {author}
  {\bibfnamefont {M.}~\bibnamefont {Wegener}},\ }\bibfield  {title} {\bibinfo
  {title} {Twisted split-ring-resonator photonic metamaterial with huge optical
  activity},\ }\href {https://doi.org/https://doi.org/10.1364/OL.35.001593}
  {\bibfield  {journal} {\bibinfo  {journal} {Opt. Lett.}\ }\textbf {\bibinfo
  {volume} {35}},\ \bibinfo {pages} {1593} (\bibinfo {year}
  {2010})}\BibitemShut {NoStop}%
\bibitem [{\citenamefont {Hopkins}\ \emph {et~al.}(2015)\citenamefont
  {Hopkins}, \citenamefont {Poddubny}, \citenamefont {Miroshnichenko},\ and\
  \citenamefont {Kivshar}}]{2015circular}%
  \BibitemOpen
  \bibfield  {author} {\bibinfo {author} {\bibfnamefont {B.}~\bibnamefont
  {Hopkins}}, \bibinfo {author} {\bibfnamefont {A.~N.}\ \bibnamefont
  {Poddubny}}, \bibinfo {author} {\bibfnamefont {A.~E.}\ \bibnamefont
  {Miroshnichenko}},\ and\ \bibinfo {author} {\bibfnamefont {Y.~S.}\
  \bibnamefont {Kivshar}},\ }\bibfield  {title} {\bibinfo {title} {Circular
  dichroism induced by fano resonances in planar chiral oligomers},\ }\href
  {https://doi.org/https://doi.org/10.1002/lpor.201500222} {\bibfield
  {journal} {\bibinfo  {journal} {Laser Photonics Rev.}\ }\textbf {\bibinfo
  {volume} {10}},\ \bibinfo {pages} {137} (\bibinfo {year} {2015})}\BibitemShut
  {NoStop}%
\bibitem [{\citenamefont {Zhen}\ \emph {et~al.}(2015)\citenamefont {Zhen},
  \citenamefont {Hsu}, \citenamefont {Igarashi}, \citenamefont {Lu},
  \citenamefont {Kaminer}, \citenamefont {Pick}, \citenamefont {Chua},
  \citenamefont {Joannopoulos},\ and\ \citenamefont {Soljačić}}]{ref63}%
  \BibitemOpen
  \bibfield  {author} {\bibinfo {author} {\bibfnamefont {B.}~\bibnamefont
  {Zhen}}, \bibinfo {author} {\bibfnamefont {C.~W.}\ \bibnamefont {Hsu}},
  \bibinfo {author} {\bibfnamefont {Y.}~\bibnamefont {Igarashi}}, \bibinfo
  {author} {\bibfnamefont {L.}~\bibnamefont {Lu}}, \bibinfo {author}
  {\bibfnamefont {I.}~\bibnamefont {Kaminer}}, \bibinfo {author} {\bibfnamefont
  {A.}~\bibnamefont {Pick}}, \bibinfo {author} {\bibfnamefont {S.-L.}\
  \bibnamefont {Chua}}, \bibinfo {author} {\bibfnamefont {J.~D.}\ \bibnamefont
  {Joannopoulos}},\ and\ \bibinfo {author} {\bibfnamefont {M.}~\bibnamefont
  {Soljačić}},\ }\bibfield  {title} {\bibinfo {title} {Spawning rings of
  exceptional points out of dirac cones},\ }\href {https://doi.org/10/f7xsjb}
  {\bibfield  {journal} {\bibinfo  {journal} {Nature (London)}\ }\textbf
  {\bibinfo {volume} {525}},\ \bibinfo {pages} {354} (\bibinfo {year}
  {2015})}\BibitemShut {NoStop}%
\bibitem [{\citenamefont {Miri}\ and\ \citenamefont {Alù}(2019)}]{ref64}%
  \BibitemOpen
  \bibfield  {author} {\bibinfo {author} {\bibfnamefont {M.-A.}\ \bibnamefont
  {Miri}}\ and\ \bibinfo {author} {\bibfnamefont {A.}~\bibnamefont {Alù}},\
  }\bibfield  {title} {\bibinfo {title} {Exceptional points in optics and
  photonics},\ }\href {https://doi.org/10/gftjt2} {\bibfield  {journal}
  {\bibinfo  {journal} {Science}\ }\textbf {\bibinfo {volume} {363}},\ \bibinfo
  {pages} {eaar7709} (\bibinfo {year} {2019})}\BibitemShut {NoStop}%
\bibitem [{\citenamefont {Lee}\ and\ \citenamefont {Chan}(2005)}]{ref65}%
  \BibitemOpen
  \bibfield  {author} {\bibinfo {author} {\bibfnamefont {J.~C.~W.}\
  \bibnamefont {Lee}}\ and\ \bibinfo {author} {\bibfnamefont {C.~T.}\
  \bibnamefont {Chan}},\ }\bibfield  {title} {\bibinfo {title} {Polarization
  gaps in spiral photonic crystals},\ }\href
  {https://doi.org/10.1364/OPEX.13.008083} {\bibfield  {journal} {\bibinfo
  {journal} {Opt. Express}\ }\textbf {\bibinfo {volume} {13}},\ \bibinfo
  {pages} {8083} (\bibinfo {year} {2005})}\BibitemShut {NoStop}%
\bibitem [{\citenamefont {Christopoulos}\ \emph {et~al.}(2019)\citenamefont
  {Christopoulos}, \citenamefont {Tsilipakos}, \citenamefont {Sinatkas},\ and\
  \citenamefont {Kriezis}}]{ref66}%
  \BibitemOpen
  \bibfield  {author} {\bibinfo {author} {\bibfnamefont {T.}~\bibnamefont
  {Christopoulos}}, \bibinfo {author} {\bibfnamefont {O.}~\bibnamefont
  {Tsilipakos}}, \bibinfo {author} {\bibfnamefont {G.}~\bibnamefont
  {Sinatkas}},\ and\ \bibinfo {author} {\bibfnamefont {E.~E.}\ \bibnamefont
  {Kriezis}},\ }\bibfield  {title} {\bibinfo {title} {On the calculation of the
  quality factor in contemporary photonic resonant structures},\ }\href
  {https://doi.org/10/gnz86x} {\bibfield  {journal} {\bibinfo  {journal} {Opt.
  Express}\ }\textbf {\bibinfo {volume} {27}},\ \bibinfo {pages} {14505}
  (\bibinfo {year} {2019})}\BibitemShut {NoStop}%
\bibitem [{\citenamefont {Verslegers}\ \emph {et~al.}(2012)\citenamefont
  {Verslegers}, \citenamefont {Yu}, \citenamefont {Ruan}, \citenamefont
  {Catrysse},\ and\ \citenamefont {Fan}}]{fanPRL}%
  \BibitemOpen
  \bibfield  {author} {\bibinfo {author} {\bibfnamefont {L.}~\bibnamefont
  {Verslegers}}, \bibinfo {author} {\bibfnamefont {Z.}~\bibnamefont {Yu}},
  \bibinfo {author} {\bibfnamefont {Z.}~\bibnamefont {Ruan}}, \bibinfo {author}
  {\bibfnamefont {P.~B.}\ \bibnamefont {Catrysse}},\ and\ \bibinfo {author}
  {\bibfnamefont {S.}~\bibnamefont {Fan}},\ }\bibfield  {title} {\bibinfo
  {title} {From electromagnetically induced transparency to superscattering
  with a single structure: A coupled-mode theory for doubly resonant
  structures},\ }\href {https://doi.org/10.1103/PhysRevLett.108.083902}
  {\bibfield  {journal} {\bibinfo  {journal} {Phys. Rev. Lett.}\ }\textbf
  {\bibinfo {volume} {108}},\ \bibinfo {pages} {083902} (\bibinfo {year}
  {2012})}\BibitemShut {NoStop}%
\bibitem [{\citenamefont {Gear}\ \emph {et~al.}(2015)\citenamefont {Gear},
  \citenamefont {Liu}, \citenamefont {Chu}, \citenamefont {Rotter},\ and\
  \citenamefont {Li}}]{JL2015PRA}%
  \BibitemOpen
  \bibfield  {author} {\bibinfo {author} {\bibfnamefont {J.}~\bibnamefont
  {Gear}}, \bibinfo {author} {\bibfnamefont {F.}~\bibnamefont {Liu}}, \bibinfo
  {author} {\bibfnamefont {S.~T.}\ \bibnamefont {Chu}}, \bibinfo {author}
  {\bibfnamefont {S.}~\bibnamefont {Rotter}},\ and\ \bibinfo {author}
  {\bibfnamefont {J.}~\bibnamefont {Li}},\ }\bibfield  {title} {\bibinfo
  {title} {Parity-time symmetry from stacking purely dielectric and magnetic
  slabs},\ }\href {https://doi.org/10.1103/PhysRevA.91.033825} {\bibfield
  {journal} {\bibinfo  {journal} {Phys. Rev. A}\ }\textbf {\bibinfo {volume}
  {91}},\ \bibinfo {pages} {033825} (\bibinfo {year} {2015})}\BibitemShut
  {NoStop}%
\end{thebibliography}%

\end{document}